\documentclass[a4paper,11pt]{article}
\pdfoutput=1
\usepackage{jheppub}
\usepackage{graphicx,amsmath,amssymb,xcolor,hyperref,slashed}
\hypersetup{colorlinks,bookmarksopen,bookmarksnumbered,
linkcolor=blus,pdfstartview=FitH,urlcolor=rossos,citecolor=verde}

\usepackage{multicol}
\usepackage{color}
\definecolor{rosso}{cmyk}{0,1,1,0.4}
\definecolor{rossos}{cmyk}{0,1,1,0.55}
\definecolor{rossoc}{cmyk}{0,1,1,0.2}
\definecolor{blu}{cmyk}{1,1,0,0.3}
\definecolor{blus}{cmyk}{1,1,0,0.6}
\definecolor{bluc}{cmyk}{1,1,0,0.1}
\definecolor{verde}{cmyk}{0.92,0,0.59,0.25}
\definecolor{verdec}{cmyk}{0.92,0,0.59,0.15}
\definecolor{verdes}{cmyk}{0.92,0,0.59,0.4}

\newcommand{\beq}{\begin{equation}}
\newcommand{\eeq}{\end{equation}}
\newcommand{\beqa}{\begin{eqnarray}}
\newcommand{\eeqa}{\end{eqnarray}}
\newcommand{\X}{S}

\title{Indirect Constraints on the Scalar Di-Photon Resonance at the LHC}

\author[1]{Florian Goertz,}
\author[1,2,3]{Jernej F. Kamenik,}
\author[1,4]{Andrey Katz,}
\author[5]{and Marco Nardecchia}

\affiliation[1]{Theory Division, CERN, 1211 Geneva 23, Switzerland}

\affiliation[2]{Jo\v zef Stefan Institute, Jamova 39, 1000 Ljubljana, Slovenia}
\affiliation[3]{Faculty of Mathematics and Physics, University of Ljubljana, \\
Jadranska 19, 1000 Ljubljana, Slovenia}
\affiliation[4]{Universit\'e de Gen\`eve, Department of Theoretical Physics and Center for Astroparticle Physics, \\24 quai E. Ansermet, CH-1211, Geneva 4, Switzerland}
\affiliation[5]{DAMPT, University of Cambridge, Wilberforce Road, Cambridge CB3 0WA, United Kingdom}

\emailAdd{florian.goertz@cern.ch}
\emailAdd{jernej.kamenik@cern.ch}
\emailAdd{andrey.katz@cern.ch}
\emailAdd{m.nardecchia@damtp.cam.ac.uk}

\preprint{CERN-PH-TH-2015-313}

\date{\today}

\abstract{Motivated by the tantalizing excesses recently reported in the di-photon invariant mass spectrum at the LHC, we scrutinize some implications of scalar di-photon resonances in high energy proton-proton collisions.  In particular, indications of a large width impose several challenges for model building. We show how calculability and unitarity considerations severely limit possible perturbative realizations  of such a signal and propose a simple criterion that can be adapted to any renormalizable model. 
Furthermore, we discuss correlations between a di-photon excess and precision observables, including the anomalous magnetic and electric dipole moments of quarks and leptons, neutral meson oscillations and radiative flavor changing neutral current mediated decays of heavy leptons and hadrons. We find that existing searches and measurements significantly constrain the possibilities for a scalar resonance decaying into final states involving Standard Model fermions. We propose future search strategies which could elucidate some remaining currently unconstrained decay channels and discuss possible correlations between the di-photon excess and several recently reported flavor anomalies,
showing that the latter can be addressed in a new incarnation of a gauged $U(1)^\prime$
model, with the di-photon resonance being the physical remnant of the $U(1)^\prime$-breaking field. 
}

\begin{document}
\maketitle

\section{Introduction}

Very recently, ATLAS reported an excess in the search for resonances decaying into photon pairs around a mass of $M_{\gamma \gamma} \sim 750$~GeV, using the first 3.2 fb$^{-1}$ of data at $\sqrt s = 13$~TeV 
collected in run II at the LHC.
The local (global) significance of the excess is 3.9\,$\sigma$ (2.3\,$\sigma$), taking the best-fit value for the width 
of the resonance of $\Gamma \sim 45$~GeV~\cite{ATLAS-CONF-NOTE}. 
Interestingly, the signal is in tentative agreement with results from the similar CMS search, which sees a local excess of 2\,$\sigma$ for the same width, employing 2.6 fb$^{-1}$ of data~\cite{CMS:2015dxe}.\footnote{For a narrow width, which seems to be preferred in CMS, the significance increases
to 2.6\,$\sigma$.} There were also less significant upward fluctuations in this mass region in the run I data collected at 8\,TeV~\cite{Aad:2015mna, CMS:2015cwa}.
The surplus might be just a statistical fluctuation, however it could also be the first hint and guidance to physics beyond the Standard Model (SM). While there have been in fact immediately numerous attempts to explain this anomaly in terms of a new particle in different concrete scenarios beyond the SM~\cite{Pilaftsis:2015ycr,
Knapen:2015dap,
Angelescu:2015uiz,
DiChiara:2015vdm,
Buttazzo:2015txu,
Backovic:2015fnp,
Harigaya:2015ezk,
Mambrini:2015wyu,
Belyaev:2015hgo,
Pelaggi:2015knk,
Dey:2015bur,
Hernandez:2015ywg,
Murphy:2015kag,
Boucenna:2015pav,
deBlas:2015hlv,
Dev:2015isx,
Chala:2015cev,
Bauer:2015boy,
Cline:2015msi,
Berthier:2015vbb,
Kim:2015ksf,
Bi:2015uqd,
Heckman:2015kqk,
Huang:2015evq,
Cao:2015twy,
Wang:2015kuj,
Antipin:2015kgh,
Han:2015qqj,
Ding:2015rxx,
Chao:2015nsm,
Barducci:2015gtd,
Cho:2015nxy,
Liao:2015tow,
Feng:2015wil,
Bardhan:2015hcr,
Chang:2015sdy,
Luo:2015yio,
Han:2015dlp,
Chang:2015bzc,
Han:2015cty,
Bernon:2015abk,
Carpenter:2015ucu,
Megias:2015ory,
Alves:2015jgx,
Gabrielli:2015dhk,
Kim:2015ron,
Benbrik:2015fyz,
Bai:2015nbs,
Falkowski:2015swt,
Csaki:2015vek,
Chakrabortty:2015hff,
Bian:2015kjt,
Curtin:2015jcv,
Fichet:2015vvy,
Chao:2015ttq,
Demidov:2015zqn,
No:2015bsn,
Becirevic:2015fmu,
Ahmed:2015uqt,
Cox:2015ckc,
Kobakhidze:2015ldh,
Cao:2015pto,
Dutta:2015wqh,
Petersson:2015mkr,
Low:2015qep,
McDermott:2015sck,
Higaki:2015jag,
Bellazzini:2015nxw,
Molinaro:2015cwg,
Patel:2015ulo,
Nakai:2015ptz,
Das:2015enc}.
Here we want to take a rather general perspective and relate the excess to other observables and considerations, which can provide consistency and phenomenological constraints on its explanations. 

Since the tentative new state (which we henceforth denote by $\X$) couples to two photons it can either have spin zero or spin two, and in the present work we will focus on the former case. Moreover, the new state can be produced singly or associated with other final states. Currently there are no experimental indications of significant additional electromagnetic or hadronic activity in signal events, and no significant missing transverse energy has been reported. All together this prefers prompt single production as the simpler explanation, which we will also concentrate on here. Theoretically, the resonance could be a singlet under the SM gauge symmetry or part of a larger electroweak multiplet. While we will mostly focus on the former case, we will comment on generalizing our results to the possibility that the state carries non-zero weak isospin. Finally, consistency of the run II excesses with run I data requires the production cross section to increase between $8$~TeV and $13$~TeV proton-proton collisions by at least a factor five, indicating production from gluon fusion or non-valence ($b,c,s$) quark annihilation~\cite{Franceschini:2015kwy}. In the present work we will focus on the first possibility, however many of our results can be easily generalized to the other cases.

As we will show in Sec.~\ref{sec:2} (and has been stressed in some of the existing literature, c.f.~\cite{Franceschini:2015kwy}) the observed signal strengths imply uncomfortably large couplings of $\X$ to photons (and gluons).  While numerous perturbative model realizations have already been proposed, they generically require the existence of large multiplicities or (color) charges of additional new massive particles (mediators), preferably close to the TeV scale. From a model building point of view,  strongly coupled extensions of the SM thus seem  to be preferred. Here we want to support and reinforce this argument showing that various supposedly renormalizable weakly coupled extensions are \textit{de facto} not calculable at all. While we cannot 
exclude that perturbation theory in any ``weakly coupled" model breaks down at the mass-scale of the mediators, we show with a concrete toy model capturing the essential ingredients of most existing proposals,  that the problem of calculability is indeed very severe.

Various authors have suggested models with couplings of order unity. Naively one could think 
that this is enough to guarantee a perturbative expansion of the theory, however, in order to fit the 
phenomenological data, a plethora of new states that carry electric charge (to generate the decay into photons) 
and color charge (in order for example to have a sizable production cross section due to the interactions with the 
gluons in the proton) have to be added to the theory. However, the large amount of new states affects the predictivity 
of the theory in two ways:  
\begin{itemize}
\item The new states modify the running of the couplings inducing the possibility of the appearance of  
Landau poles in the ultraviolet (UV). If we do not insist on having a calculable model that can be extrapolated to very high energies this is not 
a serious issue: perturbative expansion is valid, albeit  only in a limited range of energy scales;
\item The new dynamics could be such that perturbation theory breaks down already at the matching scales 
of the mediators, i.e. because the relevant expansion parameter -- the 't Hooft coupling -- grows non-perturbative. 
If this is the case any computation is not reliable. 
\end{itemize}
It is therefore important to find an easy adaptable criterion 
to understand if a given model is calculable or not. In Sec.~\ref{sec:3} we propose exactly one such criterion and 
examine it on an explicit concrete example involving new color and/or electromagnetically (EM) charged vector-like fermions with Yukawa-like couplings to $\X$. 

Once/if the $\X$ signal is established, it will also be of utmost importance to study its properties and interactions 
with SM degrees of freedom. Especially because the large decay width,
favored by the ATLAS measurements, could indicate $\X$ decay channels beyond di-photons. 
Fortunately, precision low energy measurements already provide abundant information in the form of 
constraints on the potential couplings of $\X$.  
For example, if $\X$ couples to SM quark anti-quark pairs of different generations, it will induce contributions to 
local four-Fermi interactions at low energies which can be probed with neutral meson oscillation measurements.
In addition, its sizable coupling to photons (and gluons) generically predicts the appearance of (chromo) magnetic
and electric dipole operators, which are tightly constrained by existing low energy measurements and searches. 
In particular, if the products of couplings of $\X$ to pairs of photons (gluons) and (colored) SM fermions $f$ violate
 CP, they will induce a (chromo) electric dipole moment ((C)EDM) of $f$ at the one-loop order. The CP conserving combination on the other hand generates the corresponding anomalous (chromo) magnetic dipole moment ((C)MDM). Finally, if $\X$ couples to SM fermions of different generations, it will induce radiative flavor changing neutral current (FCNC) or lepton flavor violating (LFV) transitions, both of which are already tightly constrained by  experiments.
We explore the implications of these low energy constraints on the collider phenomenology of $\X$ in 
Sec.~\ref{sec:4}.  

In few cases we also have weak hints for possible BSM physics which can also potentially be connected to the di-photon excess.   Therefore among the various observables we devote special attention to two where experimental measurements  have been exhibiting persistent tensions with SM estimates and which can be naturally linked to the existence of  a  heavy scalar resonance decaying to photon pairs: 
\begin{enumerate}
\item  The anomalous magnetic moment of the muon ($a_\mu$)
\item The measurements of rare semileptonic $B$ meson decays $B \to K^{(*)}\ell^+ \ell^-$. 
\end{enumerate}
For the former we will show that models addressing the anomaly via loop diagrams involving $S$ are in general 
tightly constrained from limits on  $S \to \mu^+ \mu^-$ decays, while for the latter we will present and constrain a simple solution with a gauged
 $U(1)^\prime$ symmetry, where the di-photon resonance is the physical remnant of the $U(1)^\prime$-breaking
field. 

At this point we note that several previous works have considered flavor constraints on the 
750~GeV diphoton resonance. In particular, Refs.~\cite{Dey:2015bur,Wang:2015kuj} have studied possible implications
of the 750~GeV resonances on the $(g-2)_\mu$ within particular models that they have analyzed. Similarly, 
Refs.~\cite{Murphy:2015kag, Bauer:2015boy} made an attempt to explain the data in conjunction with the $R_K$ and 
$R_{D^*}$ measurements, while~\cite{deBlas:2015hlv} tried to connect it to the $t \bar t$ forward-backward asymmetry.
However, we emphasize that this work is the first attempt to consider the implications of the di-photon resonance on flavor 
physics in relatively model-independent way, with generic assumptions that we outline in Sec.~\ref{sec:2}. Compared the previous works we 
also consider a much broader spectrum of flavor observables to constrain the couplings of the 
new resonance to SM fermions.

To summarize the structure of the remainder of the paper: the basic experimental facts on $S$ production and decays at the LHC are reported in Sec.~\ref{sec:2}, where we also define the theoretical framework within which our conclusions are derived. This is followed by the discussion on the calculability of weakly coupled UV realizations of the effective low energy interactions of $\X$ in Sec.~\ref{sec:3}. The constraints and implications of $\X$ couplings to SM fermions are examined in Sec.~\ref{sec:4}. We conclude with a summary of our results and discussion of possible future extensions, presented in Sec.~\ref{sec:final}.

\section{Setup}
\label{sec:2}

We follow Ref.~\cite{Franceschini:2015kwy} in parametrizing the signal strength of $S$ production through gluon fusion and its decays to pairs of photons in terms of the corresponding $S$ decay widths $\Gamma_{\gamma \gamma } \equiv \Gamma(\X\to \gamma\gamma)$ and $\Gamma_{gg } \equiv \Gamma(\X\to gg)$ computed at leading order in QCD.  Assuming $\sigma(pp\to S) \times \mathcal B (S \to \gamma\gamma) \approx (6.0\pm1.5)$~fb at $\sqrt s=13$~TeV LHC, the current data are reproduced for
\beq \frac{\Gamma_{\gamma\gamma}\Gamma_{gg}}{M\, \Gamma} \approx  8 \times 10^{-7}\,.
\label{eq:GM}\eeq
where $\Gamma\equiv \sum_f \Gamma(S\to f)$ is the total $\X$ width and $M\simeq750$~GeV its mass. In addition, the ATLAS data prefer
\beq\frac{\Gamma}{M} \approx 0.06\,.\label{eq:Gamma}\eeq  
At any scale $\mu_0$ above $M$  one can parametrize the relevant $\X$ interactions to photons and gluons in terms of local effective operators\footnote{Far above the electroweak symmetry breaking scale, one should replace the first term with its $SU(2)_L \times U(1)_Y$ invariant versions. However, this distinction has no barring on our subsequent discussion and results. }
\beq\mathcal L_F = - \frac{\epsilon_\gamma e^2 S F^2}{2 \Lambda_\gamma} - \frac{\epsilon_g g_s^2 S G^2}{2 \Lambda_g}- \frac{\tilde \epsilon_{\gamma} e^2 S F \tilde F}{2 \tilde \Lambda_{\gamma}} - \frac{\tilde \epsilon_{g} g_s^2 S G \tilde G}{2 \tilde \Lambda_{g}} \label{eq:LF}\eeq
which encode the contributions from all degrees of freedom with masses above $\mu_0$. 
We will also assume that the $\X$ field does not condense which is equivalent to requiring that the particle $\X$ is an excitation around a minimum of the scalar potential.
We can then identify $\Lambda_i$ with the mass of the lowest lying state  generating the terms in $\mathcal L_F$ and absorb additional dependence on the couplings, charges and multiplicities of such states into $\epsilon_i$.
In absence of additional contributions to $\Gamma_{\gamma\gamma}$ and $\Gamma_{gg}$  and focusing for the moment on CP even interactions, the product $(\epsilon_\gamma / \Lambda_\gamma) ({\epsilon_g}/{\Lambda_g})$ is bounded from below from~\eqref{eq:GM}, leading to\footnote{Completely analogous expressions can be derived for CP odd terms proportional to $\tilde \epsilon_{\gamma}$ and $\tilde \epsilon_{g}$ as well as possible mixed terms\,.} 
\beq
\epsilon_\gamma \epsilon_g \gtrsim 0.06 \times \frac{\Lambda_\gamma \Lambda_g}{(1\rm TeV)^2} \sqrt{\frac{\Gamma/M}{0.06}} \,,
\eeq
where the width is in general bounded from below by 
\beq
\frac{\Gamma}{M} \geq \pi \left( \alpha^2 \epsilon_\gamma^2 \frac{M^2}{\Lambda_\gamma^2}  + 8 \alpha_s^2 \epsilon_g^2 \frac{M^2}{\Lambda_g^2}  \right)\,.
\eeq
If however $\X$ couples to (in particular charged and/or colored) particles $\mathcal Q$ with masses below $\mu_0$, these have to be included in the theory. Considering for concreteness contributions due to couplings to fermions (generalization to the case of scalars or even vectors is straightforward), the relevant interactions are of the form
\beq\mathcal L_{Y} = - \sum_{\mathcal Q}  Y_{\mathcal Q} \X \bar  {\mathcal Q} P_L \mathcal Q + \rm h.c. + \ldots\,,\eeq
where $P_L \equiv (1-\gamma_5)/2$. The dots denote additional possible terms coupling different $\mathcal Q$ to $\X$, which however do not contribute to $\Gamma_{gg}$ and $\Gamma_{\gamma\gamma}$ at one loop order, as well as terms suppressed by the cut-off scale of the theory. Written in the mass basis, the sum over $\mathcal Q$ might include SM fermions. 
Then $\Gamma_{\gamma \gamma }$ and $\Gamma_{gg}$ can be written as 
\begin{align}
\frac{\Gamma_{\gamma\gamma}}{M} &= \pi \alpha^2 \left( \epsilon_\gamma^2 \frac{M^2}{\Lambda_\gamma^2}  + \tilde \epsilon_{\gamma}^2 \frac{M^2}{\tilde \Lambda_{\gamma}^2} + \left| \frac{1}{4\pi^2} \sum_{\mathcal Q} d_{\mathcal Q} Q^2_{\mathcal Q} {\rm Re}(Y_{\mathcal Q}) \mathcal F^{\rm Re}_{\mathcal Q}(2M_{\mathcal Q}/M)  \right|^2 + ( {\rm Re} \to {\rm Im} ) \right)\,, \\
\frac{\Gamma_{gg}}{M} &= 8 \pi \alpha_s^2 \left( \epsilon_g^2 \frac{M^2}{\Lambda_g^2}  + \tilde \epsilon_{g}^2 \frac{M^2}{\tilde \Lambda_{g}^2} + \left| \frac{1}{4\pi^2} \sum_{\mathcal Q} C_{\mathcal Q} {\rm Re}(Y_{\mathcal Q}) \mathcal F^{\rm Re}_{\mathcal Q}(2M_{\mathcal Q}/M)  \right|^2 + ({\rm Re} \to {\rm Im}) \right)\,, 
\end{align}
where $d_{\mathcal Q}$ is the dimension of the QCD representation of $\mathcal Q$ and $C_{\mathcal Q}$ its index while $Q_{\mathcal Q}$ is the EM charge of $\mathcal Q$. The functions $\mathcal F^{\rm Re, Im}_{\mathcal Q}(x)$ depend only on the spin of $\mathcal Q$. They are resonantly enhanced at $x \sim 1$ and decouple at large  and small $x$ as $\mathcal F^{\rm Re, Im}_{\mathcal Q}(x) \propto 1/x$ and $\mathcal F^{\rm Re, Im}_{\mathcal Q}(x) \propto x \log(x)$, respectively\,. Due to this decoupling nature, we can without loss of generality take $\mu_0 \sim M $ in which case the sums only run over SM fermions and possibly a few additional states with masses below $M$ while the rest is encoded in the coefficients $\epsilon_i$\,. All SM -- except possibly the top quark -- contributions are also completely negligible.  
We immediately see why the large required radiative widths are not easily reproduced in perturbative models. In fact, as shown in~\cite{Franceschini:2015kwy}, reproducing the data via loops of massive particles coupling to $\X$ requires either large multiplicities or large $Y$, charges and color representations. We study this issue quantitatively in the next section.

\section{Calculability in models of weakly coupled new fermions}
\label{sec:3}

To start our discussion, we will introduce a criterion to check if a given model is calculable.
Assuming that a model contains a series of couplings $y_i$, then, inspired by the 1/$16 \pi^2$ suppression in the loop expansion, one could suggest that a plausible criterion is the requirement that $|y_i | < 4 \pi$ for every coupling in the theory. However, considering the New Physics models that aim at explaining the di-photon resonance, this proposal could be misleading. If there are $N$ copies of mediators propagating in the loop, it is clear that the multiplicity of the states has to be considered in the criterion. For example the requirement 
$\left| y_i \right| < \frac{4 \pi }{\sqrt{N}}$, \`a la 't Hooft,  could be more meaningful than $|y_i | < 4 \pi$. 

We can try to improve the definition of our criterion in the following way. The value of the coupling $y_i$ depends on the energy scale $\mu$ and its rate of change with energy is dictated by the beta function $\beta_{y_i}$. If we change the energy scale by an amount $\delta \mu$ we get a variation to the coupling $y_i$ that amounts to 
\begin{equation}
\delta y_i  \approx \frac{\partial y_i }{\partial \mu} \, \delta \mu 
\end{equation}
The relative variation is given by
\begin{equation}
\frac{\delta y_i }{y_i} \approx \frac{\mu}{y_i} \frac{\partial y_i }{\partial \mu} \, \frac{ \delta \mu }{ \mu} = \frac{\beta_{y_i}}{ y_i} \frac{ \delta\mu }{ \mu} 
\end{equation}
We conclude that if the the ratio $|\beta_{y_i} /  y_i |$ is large, the coupling constant varies a lot even with small changes of the renomalization scale and henceforth one runs into troubles with calculability.
Based on this, our final proposal for the criterion is
\begin{equation}
\label{crit}
\left| \frac{ \beta^{(1)}_{y_i} ( \{ y \} )}{y_i} \right| < 1 \qquad \textrm{for every coupling $y_i$}\,,
\end{equation}
where $\beta^{(1)}_{y_i}$ is the $\beta$ function of the coupling $y_i$ truncated at the one-loop level. The $\beta$ functions are obtained as the sum of various contributions, so avoiding accidental cancellation, we will apply the proposed bound (\ref{crit}) to every contribution. This criterion benefits also from the fact that the one-loop beta functions for renormalizable gauge-Yukawa theories are well known and easy to compute (actually complete formulae are available up to two loops~\cite{Machacek:1983tz,Machacek:1983fi,Machacek:1984zw,Luo:2002ti}). 

We will now analyze the restriction imposed by (\ref{crit}) on a simple toy model. Despite its simplicity, our toy model captures the main features of several proposals which have appeared in the literature.
Only in this section, we use a two-component notation for fermions.
We introduce $N_Q$ copies of neutral vector-like QCD triplets $(Q_A,Q^c_A )$ as well as $N_E$ copies of colorless vector-like fermions $(E_B,E^c_B )$, singlet under $SU(2)_L$ and with hypercharge $Y$. We assume the theory to be invariant under a $SU(N_Q) \times SU(N_E)$ global symmetry.
The 750 GeV resonance is represented by a real pseudo-scalar field $S$.
We also impose invariance under parity transformation. In two-component notation the transformation rules that we assume are
\begin{eqnarray*}
Q_A \left(\mathbf{x}, t\right) & \to & \left( Q^{c}_A \left(- \mathbf{x}, t\right) \right)^{\dagger}\,, \\
E_B \left(\mathbf{x}, t\right) & \to & \left( E^{c}_B \left(- \mathbf{x}, t\right) \right)^{\dagger}\,, \\
S \left(\mathbf{x}, t\right) & \to & - S \left(-\mathbf{x}, t\right)\,.
\end{eqnarray*}
The most general Lagrangian beyond the SM with the above assumed symmetries is given by
\begin{eqnarray*}
\mathcal{L_{\textrm{NP}}} &=& i Q^{\dagger}_A \sigma^{\mu}D_{\mu} Q_A +i Q^{c \dagger}_A \sigma^{\mu}D_{\mu} Q^c_A 
+i L^{\dagger}_B \sigma^{\mu}D_{\mu} L_B +i L^{c \dagger}_B \sigma^{\mu}D_{\mu} L^c_B \\
& -& \left( M_Q Q^c_A Q_A + M_E E^c_B E_B + i y_q Q^c_A Q_A S + i y_e E^c_B E_B S + \textrm{h.c.} \right) \\
& -& \left( \frac{M^2}{2} S^2 + \frac{\lambda}{4!} S^4 \right)\, .
\end{eqnarray*}
Invariance under parity forces $y_q$ and $y_e$ to be real and it is responsible also for the absence of linear and cubic terms in $S$ in the potential.
We are also omitting for simplicity the scalar quartic coupling $H^\dagger H S^2$.

In order to apply our criterion, we need to compute one-loop $\beta$ functions of the couplings $g_s, g', y_q, y_e$ and $\lambda$.
Using the analytic expressions from \cite{Luo:2002ti}, we obtain:
\begin{eqnarray}
\beta_{g_s} & = &  \frac{g^3_s}{16 \pi^2}  \left( 7 - \frac{2}{3} N_Q \right),\\
\beta_{g'} & = &   \frac{g'^3}{16 \pi^2} \left(\frac{41}{6} + \frac{4}{3}N_E Y^2 \right), \\
\beta_{y_q} &=& \frac{y_q }{16 \pi^2} \left( 3 y_q^2 + 2  \left(3 N_Q y_q^2 + N_E y_e^2 \right) - 8 g_s^2 \right), \\
\beta_{y_e} &=& \frac{y_e }{16 \pi^2} \left( 3 y_e^2 + 2  \left(3 N_Q y_q^2 + N_E y_e^2 \right) - \frac{18}{5} g'^2 Y^2 \right), \\
\label{betalam}
\beta_{\lambda} & =& \frac{1}{16 \pi^2} \left[ 3 \lambda^2 + 8  \left( 3 N_Q y_q^2 + N_E y_e^2 \right) \lambda - 48  \left( 3 N_Q y_q^4 + N_E y_e^4 \right)   \right]\,.
\end{eqnarray}
From the $\beta$ functions of the gauge couplings we get:
\begin{eqnarray}
\left| \frac{\beta_{g_s}}{g_s} \right| < 1  & \Longrightarrow & N_Q < \frac{3}{2} \left( \frac{4 \pi}{\alpha_s} + 7\right) \approx 200\,, \\
\left| \frac{\beta_{g'}}{g'} \right| < 1  & \Longrightarrow & N_E Y^2<  \frac{3}{4} \left( \frac{4 \pi}{\alpha'} -\frac{41}{6} \right)  \approx 1580\,.
\end{eqnarray}
The numerical approximations are obtained assuming $\alpha_s = 0.1$ and $\alpha=1/128$. These bounds on $N_E$ and $N_Q$ are very loose; however, our criterion can be extended to every monomial of every $\beta$ function. We crucially observe that the size of $\lambda$ is not protected by any symmetry. Indeed, in the limit $\lambda \to 0$ there is no enhancement of the global symmetry. This also means that $\lambda$ is the most favorable coupling to get very large radiative corrections, so we expect $\left| \beta_{\lambda}/\lambda \right|<1$ to give important constraints.
In particular, considering the various terms in (\ref{betalam}), we derive the following bounds:
\begin{eqnarray}
\frac{3}{16 \pi^2} \left| \lambda \right| &<& 1\,, \\
\frac{3}{2 \pi^2} N_Q y_q^2 &<& 1\,, \\
\frac{1}{2 \pi^2} N_E y_e^2 &<& 1\,, \\
\label{yqquart}
\frac{9}{\pi^2} N_Q y_q^4 &<& \left| \lambda  \right| , \\
\label{yequart}
\frac{3}{\pi^2} N_E y_e^4 &<& \left|  \lambda \right|  .
\end{eqnarray}
The bound on $\lambda$ derived in this way is again very loose but we can find a maximum allowed value for this parameter imposing the condition of non-violation of perturbative unitary for the scattering of scalars $SS \to SS$. The analysis of the s-wave at tree-level gives $ |\lambda| < 16 \pi $. 
Making the conservative assumption $|\lambda|\leq 16 \pi$, from (\ref{yqquart}) and  (\ref{yequart}), we get:
\begin{equation}
N_Q y_q^4 < 55.1\,, \qquad N_E y_e^4 < 165 \,.
\end{equation}
It is possible to extract other bounds from $\left| \frac{\beta_{y_q}}{y_q} \right| < 1$ and $\left| \frac{\beta_{y_e}}{y_e} \right| < 1$ but it turns out that the constraints derived from $\left| \beta_{\lambda}/\lambda \right| < 1 $ are always stronger in any region of the parameter space.
We are now ready to compare these bounds with the information on the model parameters coming from data on $S$.
To this end, we can use the expression of Section 3 of \cite{Franceschini:2015kwy} adapted to the case of CP-odd interactions.
The induced widths from fermion loops are given by
\begin{eqnarray}
\Gamma(\X\to gg) &=& M \frac{\alpha_3^2}{8\pi^3} N^2_Q y^2_q \tau_Q \left| \mathcal P(\tau_Q) \right|^2\,, \\
\Gamma(\X\to \gamma\gamma) &=& M \frac{\alpha^2}{16\pi^3} Y^4 N^2_E y^2_e \tau_E \left| \mathcal P(\tau_E) \right|^2\,,
\end{eqnarray}
where  $\tau_Q = 4M_Q^2/M^2$ and $\tau_E = 4M_E^2/M^2$ and the loop function is defined as
\begin{equation}
\mathcal P(\tau)   = \arctan^2(1/\sqrt{\tau-1})\,.
\end{equation}
In order to be conservative, we take the values of mediator masses close to their expected experimental exclusion limit. In particular, we take $M_Q = 1 $ TeV and $M_E = 400 $ GeV. The decay widths normalized to the mass of the scalar are given by
\begin{equation}
\frac{\Gamma \left(S \to gg \right)}{M} =5.7 \cdot 10^{-6} \, y^2_q N^2_Q\,, \qquad \frac{\Gamma \left( S \to \gamma \gamma \right)}{M} =  1.1 \cdot 10^{-7} \, Y^4 y^2_e N^2_E\,,
\end{equation}
and we neglected corrections of order $\left(M / 2 M_Q\right)^2$  and $\left(M / 2 M_E\right)^2$, respectively.
Imposing the bounds on the product of $\Gamma_{\gamma \gamma} \Gamma_{gg}$, we obtain
\begin{equation}
Y^4 N^2_E N^2_Q y_e^2 y_q^2 = 7.2 \cdot 10^4\,.
\end{equation}

Extra constraints can be derived considering other phenomenological aspects. The non-observation of any significant excess in the di-photon invariant mass distribution at  the 8 TeV run at the LHC, suggests that the production cross section increases sizeably when varying the energy from $\sqrt{s} = 8$ TeV to $\sqrt{s} = 13$ TeV.  This fact favours production mechanisms of the scalar $S$ with large gain factor $r \equiv \sigma_{\textrm{13 TeV}} / \sigma_{\textrm{8 TeV}} $. In our toy model, the scalar $S$ can be produced by gluon or photon fusion, and the respective gain factors are given by $r_{\gamma \gamma} =Ê1.9$ and $r_{gg}=4.7$.  Henceforth, to have a better fit of the 8 TeV and 13 TeV data, we impose in our model that the gluon production dominates over the photon one: 
\begin{equation}
C_{gg} \Gamma_{gg} > C_{\gamma \gamma} \Gamma_{\gamma \gamma}
\end{equation}
where $C_{gg}=2137$ and  $C_{\gamma \gamma}=54$ are the partonic integrals as defined in~\cite{Franceschini:2015kwy}.
In terms of our parameters we get
\begin{equation}
y_q^2 N_Q^2 > 4.9 \cdot 10^{-4} \,  Y^4 y_e^2 N_E^2 \, .
\end{equation}

An upper bound on $\Gamma_{gg}$ can be derived using results for searches for resonances decaying to di-jet final states~\cite{Aad:2014aqa}. From the analysis of~\cite{Franceschini:2015kwy} we infer that $\Gamma_{gg} / M < 2 \cdot 10^{-3}$ and this gives:
\begin{equation}
y^2_q N^2_Q < 2.6 \cdot 10^{5}.
\end{equation}

We are now ready to collect all the information and to check in which region of the parameter space the model can be considered calculable according to our criterion (\ref{crit}).
We obtain
\begin{equation}
\label{system}
\begin{cases}
N_Q < 200  \qquad &\textrm{from $| \beta_{g_s}/ g_s | < 1 $}\,, \\
N_E Y^2 < 1580  \qquad & \textrm{from $| \beta_{g'}/ g' | < 1 $}\,,  \\
N_Q y_q^2 < 6.6 \qquad  & \textrm{from $| \beta_{\lambda}/ \lambda | < 1 $}\,, \\
N_E y_e^2 < 19.7   \qquad & \textrm{from $| \beta_{\lambda}/ \lambda | < 1 $}\,, \\
N_Q y_q^4 < 55.1 \qquad  & \textrm{from $| \beta_{\lambda}/ 16 \pi | < 1 $} \,,\\
N_E y_e^4 < 165   \qquad & \textrm{from $| \beta_{\lambda}/ 16 \pi | < 1 $}\,, \\
Y^2 N_E N_Q y_e y_q = 268 & \textrm{$\sigma ( pp \to S \to \gamma \gamma )=8$ fb}\,, \\
Y^2 \left| \frac{y_e }{y_q} \right| \frac{N_E}{N_Q} < 45 & \textrm{$C_{gg} \Gamma_{gg} > C_{\gamma \gamma} \Gamma_{\gamma \gamma}$}\,, \\
|y_q| N_Q < 513 & \Gamma_{gg}/M < 2 \cdot 10^{-3}\,, \\
\end{cases}
\end{equation}
and will use the numerical analysis of this model to draw conclusions that are rather general. Moreover, we will comment on the range of applicability of our analysis and, finally, will provide the $\beta$ function for a generic gauge-Yukawa model with a single scalar field.

The analysis of the system of constraints in (\ref{system}) suggests two scenarios where the model is calculable at the threshold of new physics states:
\begin{enumerate}
\item {\it Large hypercharge scenario.}

If the hypercharge of the new states is very large, then all of our constraints can be easily satisfied. The qualitative reason is again linked to the radiative correction of the self-interaction $\lambda$. Indeed, if we change the value of $Y$, all the constraints generated by $ |\beta_{\lambda}  / \lambda| <1 $ remain unaffected, while the contribution to $\Gamma_{\gamma \gamma}$ gets a sufficient enhancement to explain the properties of the 750 GeV resonance (including the large decay width). 
Notice that a change in the $SU(3)_C$ color irreducible representation can be more dangerous because even if we benefit from an enhancement in $\Gamma_{gg}$ due to larger group invariant $C_r$, the enlargement of the dimensionality of the irreducible representation gives a larger correction to the quartic $\lambda$.

\item {\it Large multiplicity scenario.}

If we want to avoid very large and exotic electric charges, then we are forced to consider a sizable number of states. In the following we will consider $Y=1$ for concreteness to present the argument.
First, just considering the seventh row of (\ref{system}), it seems possible to address the excess by assuming Yukawa couplings of $y_e \sim y_q \sim 1$ with $N_E \sim N_Q \sim 16$ or even 
$y_e \sim y_q \sim {\cal O}(5)$ with only $N_E \sim N_Q \sim 3$.
However, taking into account in particular the third and fourth rows, such solutions are excluded.
So we make the crucial observation that the running of the scalar quartic renders scenarios trying to address the signal with 
sizable Yukawa couplings and a modest number of new states uncalculable. A very large number of new states seems 
unavoidable. Quantitatively, from the third, fourth, and seventh rows in (\ref{system})
above, we obtain the interesting bounds
\begin{equation}
\label{eq:NENQ}
N_E N_Q > 550\,,
\end{equation}
and
\begin{equation}
y_e y_q < 0.49 \,.
\end{equation}

Finally, we present a quantitative overview on the number of required new fermions and its interplay with the Yukawa couplings of the new scalar in Figure \ref{fig:NN}. 
Here, we show contours of constant Yukawa couplings $y_e=y_q$, which we assume
equal for simplicity, in the $N_E$ vs.\ $N_Q$ plane, where we require the seventh row of (\ref{system}) to be fulfilled. 
The allowed region is obtained for
\begin{equation}
41 < N_E < 1580\,, \qquad 14 < N_Q < 200 \, .
\end{equation}
	\begin{figure}[!t]
	\begin{center}
	\includegraphics[height=3.12in]{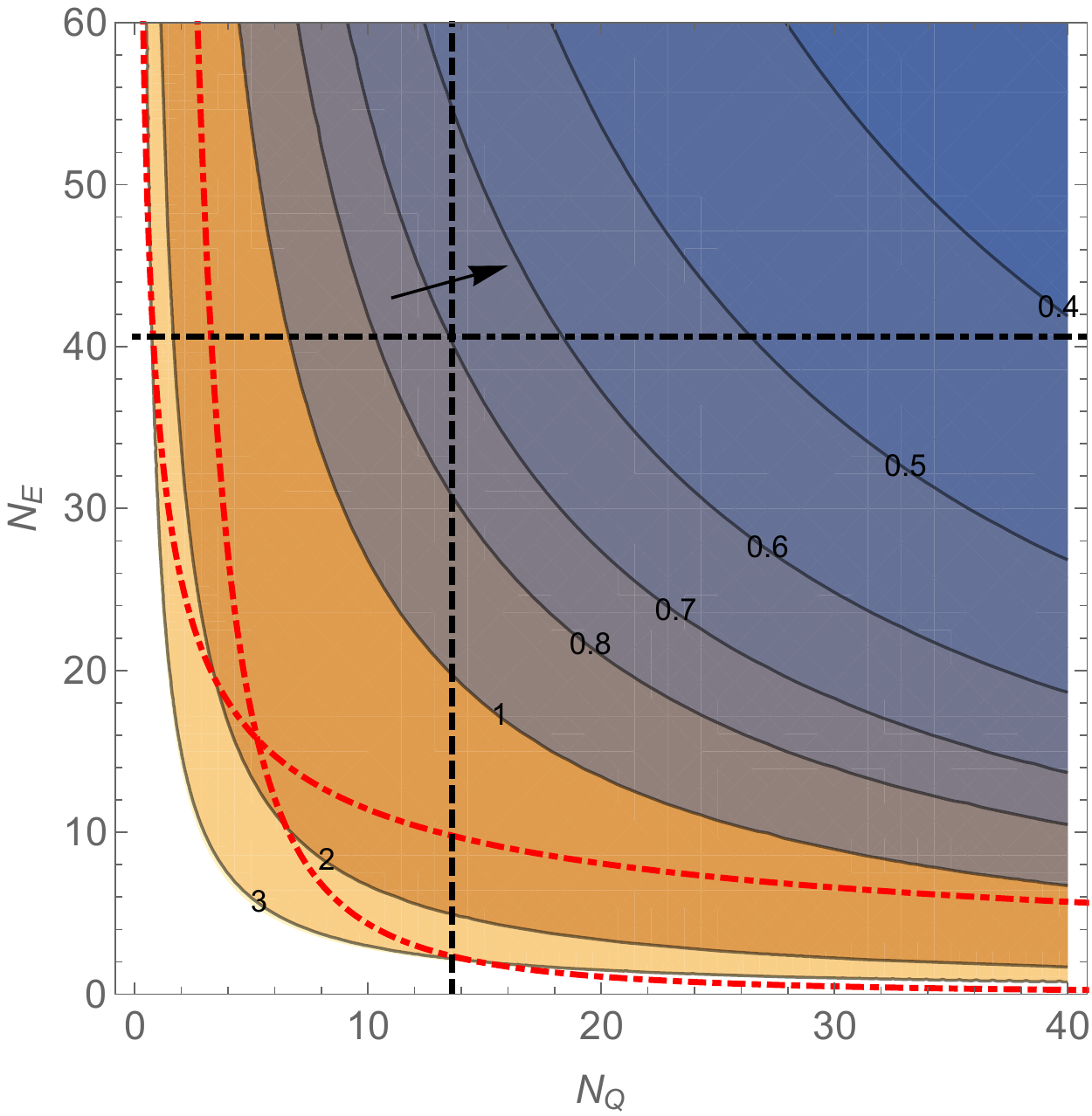}
	\caption{\label{fig:NN} Contours of constant Yukawa couplings $y_e=y_q$ in the $N_E$ vs. $N_Q$ plane for parameter points 
reproducing the di-photon resonance. 
The bounds from the running of the scalar quartic from the third and forth rows of (\ref{system}) are depicted by black dashed-dotted lines,
while bounds from the fifth and sixth rows are depicted by red dashed-dotted lines. The arrow denotes the region of parameters were all constraints are satisfied. See text for details.}
	\end{center}
	\end{figure}	
Although the model can be considered calculable according to~(\ref{crit}), the vary large number of states generates a Landau pole very close to the TeV scale.
\end{enumerate}


The considerations made for our toy model are rather general and we believe that the problem of calculability is an issue in a large part of the proposed weakly coupled solutions of the 750 GeV puzzle. For this reason, we suggest applying our simple strategy in order to check if a model is calculable and predictions can be trusted. We summarize here the main considerations to perform this analysis:
\begin{itemize}
\item Our criterion is based on a comparison between tree-level effects and one-loop corrections. The comparison is done confronting the tree-level value of the couplings with their $\beta$ function:
\begin{equation}
\label{crit2}
\left| \frac{ \beta^{(1)}_{y_i} ( \{ y \} )}{y_i} \right| < 1 \qquad \textrm{for every coupling $y_i$}\,.
\end{equation}
\item One should write the complete sets of beta functions for a given model, these can be found for example in \cite{Machacek:1983tz,Machacek:1983fi,Machacek:1984zw,Luo:2002ti}.
\item If we don't allow cancellations of the various terms in the $\beta$ functions, our condition can be applied to every monomial .
\item This will give a large number of constraints on the parameters space, and one has to check if this region allows for an explanation of the 750 GeV resonance.
\item We also want to remark that a quartic coupling for the new scalar state is always allowed in a renormalizable model. Moreover, this term is not protected by any symmetry, so it is the ideal place to look for large radiative corrections.\footnote{With few exceptions: $D$-terms in SUSY extensions, where quartic scalars are protected by gauge symmetry and the goldstone boson case, where the protection is given by the shift symmetry.} 
\item Perturbative unitarity bounds can be used to set upper bounds on the couplings providing further constrains; for example, those coming from  $2 \to 2$ scattering of the new scalar are particularly important to bound the quartic self-coupling.
\end{itemize}

We conclude by noticing that several explanations of the 750 GeV resonance consider just the SM augmented by a single real scalar and a series of fermions. For this case, we provide simple and compact formula for the $\beta$ functions based on the results in \cite{Luo:2002ti}. We consider the following lagrangian for a simple gauge group
\begin{eqnarray}
\mathcal L & = & - \frac{1}{4} F^{A \mu \nu}  F^A_{\mu \nu} + \frac{1}{2} \partial^{\mu} S \partial_{\mu} S + i \psi^{\dagger}_{i} \sigma^{\mu} D_{\mu} \psi_i \\
& &- \frac{1}{2} \left(\underline{Y}_{ij} \psi_i \psi_j S + \textrm{h.c.} \right) - \frac{1}{4!} S^4\,.
\end{eqnarray}
Lorentz contractions are implicit and we consider only marginal operators. The matrix $\underline{Y}$ is such that it extends over all the indices of the fields including the gauge ones. For example, for a pair of vector-like QCD triplets $Q$ and $Q^c$, $\underline{Y}$ is symmetric matrix of order 6. We also need the following definitions:
\begin{eqnarray}
S_2 (F) \delta^{AB} & = & \textrm{Tr}\left(t^A t^B \right) \,,\\
C_2 (G) \delta^{AB} & = & f^{ACD} f^{BCD} \,,\\
C^{ab}_2 (F) & = & t^A_{ac} t^A_{cb}\,,
\end{eqnarray}
where $t^A$ are the generators of the gauge group acting on the fermion fields and $f^{ABC}$ are the structure constant of the simple group. The capital letter $F$ refers to the (in general reducible) representation of the fermions under $G$ denote the adjoint irreducible representation of the gauge group.
The $\beta$ functions are given by:
\begin{eqnarray}
\beta_g & = &- \frac{g^3}{(4 \pi)^2} \left(\frac{11}{3} C_2(G) -\frac{4}{3} \kappa S_2 (F) \right)\,, \\
\beta_{\lambda}  &=& \frac{1}{16 \pi^2} \left(3 \lambda^2 + 8 \kappa \, \textrm{Tr} \left( \underline{Y}^{\dagger} \underline{Y} \right) \lambda - 48 \kappa \, \textrm{Tr} \left( (\underline{Y}^{\dagger} \underline{Y})^2 \right)  \right) \,,\\
\beta_{\underline{Y}} &=& \frac{1}{2} \underline{Y}^{\dagger} \underline{Y}^2 + \frac{5}{2} \underline{Y}  \underline{Y}^{\dagger} \underline{Y} + 2 \kappa \, \underline{Y} \textrm{Tr} \left( \underline{Y}^{\dagger} \underline{Y}\right)\,,
-3 g^2 \{   C_2(F), \underline{Y} \}  
\end{eqnarray}
where the parameter $\kappa$ is set to 1/2 for the case of two-component fermions. 
An extension to non-simple groups $G_1 \times \dots \times G_n$ requires the substitutions:
\begin{eqnarray}
g^3 C_2(G) &\to& \sum_k g_k^3 C_2(G_k)\,, \\
g^3 S_2(F) &\to& \sum_k g_k^3 S^k_2(F)\,, \\
g^2 C_2(F) &\to& \sum_k g^2_k C^k_2(F)\,,
\end{eqnarray}
where $g_k$ is the gauge coupling of the group $G_k$. 
As we argued in this section, we expect the most important constraints to came from the radiative effects on $\lambda$. In particular we get 
\begin{eqnarray}
\left| \frac{\beta_{\lambda}}{\lambda} \right| < 1  & \Longrightarrow &  \textrm{Tr} \left( \underline{Y}^{\dagger} \underline{Y} \right) < 4 \pi^2 \approx 39\,,\\
\left| \frac{\beta_{\lambda}}{16 \pi} \right| < 1  & \Longrightarrow & \textrm{Tr}  \left( (\underline{Y}^{\dagger} \underline{Y})^2 \right)  <  \frac{2 \pi^2}{3} 16 \pi  \approx 330\,.
\end{eqnarray}

\section{Constraints on couplings to SM fermions}
\label{sec:4}

As discussed in Sec.~\ref{sec:2}, we can parametrize the relevant $\X$ couplings to SM fermions after electroweak (EW) symmetry breaking -- in the mass basis -- as
\beq
\mathcal L_{Y} \ni - \sum_{\ell_i,\ell_j = e,\mu,\tau}  Y_{\ell_i\ell_j} \X \bar \ell_i P_L \ell_j  -  \sum_{d,d_j = d,s,b}  Y_{d_i d_j} \X \bar d_i P_L d_j  -  \sum_{u_i,u_j = u,c,t}  Y_{u_i u_j} \X \bar u_i P_L u_j + \rm h.c.\,.
\eeq
In the following we also fix $\Lambda_i$ = 1 TeV in~\eqref{eq:LF}. All our results however can be easily rescaled to any other values. Considering effects well below the scale of $M$, we can also safely use~\eqref{eq:LF} to parametrize the constraints coming from the observations of $S$ at the LHC.
The summary of our derived constraints on $Y_{ij}$ and consequently $\X$ decay products with these assumptions is given in Table~\ref{tab:1}. The details of their derivation are however presented in the following two subsections.
\begin{table}
\begin{center}
\begin{tabular}{|c|c|c|}
\hline
Bound on $Y_{f,f'}$ & Observable & $\Gamma(\X\to f f')/M$ \\
\hline
\hline
$|{\rm Im}(Y_{ee})| \lesssim 1 \times 10^{-7}$ & $d_e $& $ \lesssim 6 \times 10^{-16}$ \\
\hline
$|{\rm Im}(Y_{dd})| \lesssim 3 \times 10^{-4}$ & $d_N, d_{\rm Hg} $ & $ \lesssim 1 \times 10^{-8}$ \\
$|{\rm Im}(Y_{uu})| \lesssim 5  \times 10^{-4}$ & $d_N, d_{\rm Hg} $ & $ \lesssim 3 \times 10^{-8}$ \\
\hline
$|Y_{e\mu}|, |Y_{\mu e}| \lesssim 2 \times 10^{-5}$ & $\mathcal B(\mu \to e \gamma) $& $\lesssim 1 \times 10^{-11}$ \\
$|Y_{e\tau}|, |Y_{\tau e}| \lesssim  0.08$ & $\mathcal B(\tau \to e \gamma)$ & $\lesssim  3 \times 10^{-4}$ \\
$|Y_{\mu\tau}|,|Y_{\tau\mu}| \lesssim 0.09$ & $\mathcal B(\tau \to \mu \gamma)$ & $\lesssim 3 \times 10^{-4}$ \\
\hline
$\sqrt{{\rm Re}[(Y_{sd})^2]}, \sqrt{{\rm Re}[(Y_{ds})^2]}  < 1.3 \times 10^{-4}$ & $ \Delta m_K$ & $<2.0 \times 10^{-9}$ \\
$\sqrt{{\rm Im}[(Y_{sd})^2]}, \sqrt{{\rm Im}[(Y_{ds})^2]} < 3.4 \times 10^{-6}$ & $\epsilon_K$ & $<1.4 \times 10^{-12}$ \\
$\sqrt{{\rm Re}[(Y_{cu})^2]}, \sqrt{{\rm Re}[(Y_{uc})^2]}< 3.3 \times 10^{-4}$ & $ x_D $ & $<1.3 \times 10^{-8}$ \\
$\sqrt{{\rm Im}[(Y_{cu})^2]}, \sqrt{{\rm Im}[(Y_{uc})^2]}< 4.0 \times 10^{-5}$ & $ (q/p)_D, \phi_D $ & $<1.9 \times 10^{-10}$ \\
$\sqrt{{\rm Re}[(Y_{bd})^2]}, \sqrt{{\rm Re}[(Y_{bd})^2]}< 4.1 \times 10^{-4}$ & $ \Delta m_d $ & $<2.0 \times 10^{-8}$ \\
$\sqrt{{\rm Im}[(Y_{bd})^2]}, \sqrt{{\rm Im}[(Y_{bd})^2]}< 2.3 \times 10^{-4}$ & $ \sin2\beta $ & $<6.3 \times 10^{-9}$ \\
$|(Y_{bs})|, |(Y_{sb}) | < 1.7\times 10^{-3}$ & $ \Delta m_s $ & $<3.4 \times 10^{-7}$ \\
\hline
\end{tabular}
\end{center}
\caption{\label{tab:1} Constraints on $\X$ couplings to SM fermions (first column) derived from low energy precision observables (second column). Results in first seven rows assume CP-even $S$ ($\tilde \epsilon_{g}=\tilde \epsilon_{\gamma}=0$), $|\epsilon_\gamma| > 0.05 $ and $|\epsilon_g|>1 \times 10^{-3}$ in order to reproduce the observed di-photon signal from gluon fusion. The case of CP-odd $S$ can be obtained via replacement of ${\rm Im} \to {\rm Re}$.  Only a single $Y_{ij}$ is assumed to be non vanishing in each column. The associated limits on the $\X$ decay rates to  the corresponding flavored final states are presented in the third column. }
\end{table}

\subsection{Dipole moments}
Important information and constraints on the 
flavor and CP structure of $S$ couplings to SM fermions can be extracted from existing low energy measurements and searches. In particular, focusing on the flavor diagonal couplings, 
the measured MDMs of SM charged leptons, the constraints on their EDMs, as well as constraints on nuclear EDMs provide very stringent probes. In order to simplify the notation in this section we will explicitly split each flavor diagonal $S$ Yukawa
coupling into its real and imaginary parts as
\beq
Y_{ii} \equiv Y'_{i}+ i \tilde Y'_{i}~, 
\eeq
where $Y'_i$ and $\tilde Y'_i$ are defined as real. We start by defining the NP induced SM lepton EDMs and shifts to MDMs via
\beq
\mathcal L^\ell_{\rm dipole} = - \sum_\ell \frac{1}{4 m_\ell} \bar \ell \sigma^{\mu\nu} \left( e \,\Delta a_\ell + 2 i\, m_\ell \,d_\ell \,\gamma_5 \right) \ell F_{\mu\nu} \,.
\eeq
We first analyze the constraints (as well as possible NP hints) coming from the lepton MDM 
measurements. The current experimental situation can be summarized as follows~\cite{Hanneke:2010au,Abdallah:2003xd, Agashe:2014kda}
\beqa
|\Delta a_e|  & < & 8 \times 10^{-12}\,,
\\ \label{eq:muMDM}
\Delta a_\mu & = &  (2.92 \pm 0.86) \times 10^{-9}\,, \\
-0.052 & < & \Delta a_\tau < 0.013 \, ,
\eeqa
where in the second row we have added the various experimental and theoretical errors in quadrature and took the central value as quoted in the 
PDG~\cite{Agashe:2014kda}. 
 Perhaps the most interesting part here is the long standing anomaly in $a_\mu$. 
 If explained due to the same new particles, which are responsible for the $\gamma \gamma $ excess at 750~GeV, 
 it can provide unambiguous information on the couplings of the new sector to muons and further correlate $a_\mu$ with existing constraints on the  $S$ decays to $\mu^+ \mu^-$. 
 
\begin{figure}[!t]
\begin{center}
\includegraphics[height=2.12in]{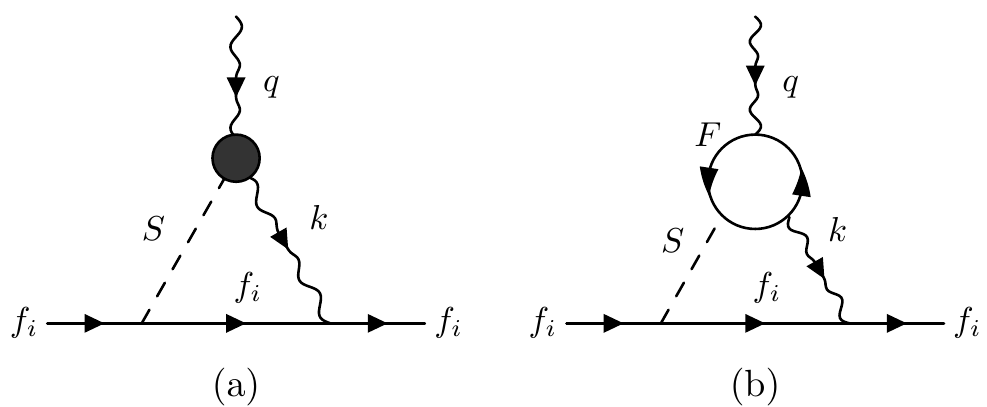}
\caption{\label{fig:BarrZee} Barr-Zee diagrams which contribute to the EDMs and MDMs or the SM fermions. The diagram on the RH side shows the full perturbative two-loop process, while in the diagram on the LH-side we replace 
the $S\gamma \gamma $ interaction by an effective vertex. }
\end{center}
\end{figure}

The dominant contribution to the muon MDM comes from 
 the so called Barr-Zee type diagram in Fig.~\ref{fig:BarrZee}, very similar to the structure 
 shown in~\cite{Chang:2000ii,Dedes:2001nx, Cheung:2001hz,Cheung:2003pw,Gunion:2008dg} 
 in the context of the Two Higgs doublet model (THDM).\footnote{ We note in passing that while focusing here on an EW singlet $S$, our results are expected to apply also to large parts of the parameter space of models where $S$ is the neutral component of a larger multiplet.  The main difference is due to the fact that in the singlet case there is 
 only as single neutral state which gives a new contribution to the MDM. In a THDM on the other hand
 one in general needs to include both CP-even and CP-odd states. 
 Moreover, these contributions are suppressed only by $\log M$
 and therefore heavier states are potentially as important as of the lightest one. Hereafter we assume only one 
 state in our analysis, but extension to two or more states is possible along these lines.}
 There are however several important differences between the perturbative THDM scenarios and the required properties of $S$.  First, as discussed in the previous section the substantial coupling of $S$ to the di-photons is likely  to be explained within a non-perturbative UV completion. 
In accordance with this and in order to simplify the expressions, we will also completely neglect the SM fermions running in the loop and 
 concentrate on the UV contributions in~\eqref{eq:LF}.  Second, 
 the likely non-perturbativity of the UV completion  poses a clear 
 challenge to the explicit calculation of the Barr-Zee diagrams when the effective couplings between  $S$ and photons emerges from strong dynamics. 
 To emeliorate this problem, we consider in our analysis only the universal pieces, logarithmically sensitive to the UV dynamics which scale as $ \log (\Lambda_i)/\Lambda_i$, but neglect all the finite terms which scale as $1/\Lambda_i$, because the latter are model dependent.  

Before we go into the details of the calculation, it is instructive to estimate the effect from considerations of dimensional analysis. Being a logarithmically-divergent contribution to a dimension-five operator, it is expected to scale as 
 \beq
 \Delta a_\ell \sim \epsilon_\gamma 
 \frac{m_\ell}{\Lambda_\gamma} \frac{\alpha\, Y'_\ell}{4 \pi} \log \frac{\Lambda_\gamma^2}{M^2} + 
 {\rm finite}~,
 \eeq
 where $\Lambda_\gamma$ is the mass scale of heavy (strongly coupled) states generating $S\to \gamma\gamma$.
 This structure immediately suggests that accommodating the reported value
 of the $a_\mu$ anomaly~\eqref{eq:muMDM}
 requires $Y'_\mu \sim 0.1$, in conflict with the
 non-observation of $S$ in the $\mu^+ \mu^-$ channel. 
 We will now show that this naive estimation is in fact 
 fairly close to the result of the explicit calculation.

In our estimate we assume that the structure of the $S\gamma \gamma$ vertex is 
 identical to one gets from loops of heavy fermions (see~\cite{Ilisie:2015tra}
 for the details), namely
 \beq
 i\Gamma^{\mu\nu} \propto i \Lambda (g^{\mu \nu} k \cdot q - k^\mu q^\nu) \int_0^1 dx 
 \frac{2 x(1-x) - 1}{k^2(1-x) - \Lambda^2}, 
 \eeq  
 where $q^\mu$ stands for the momentum of the external photon, $k^\mu$ denotes the momentum running in the loop, 
 and $\Lambda$ is the mass of the heavy fermion running in the loop. Of course, at the end of the day we will take the 
 limit $\Lambda \to \infty $ to recover the effective dimension-5 operator. The first part of this expression is completely generic and relies on no model assumptions. It is only the integral which depends on the structure generated by integrating out heavy fermions. However, we expect these details of the vertex structure to be relevant only for the finite terms, and not for the leading logarithm that we are interested in. 

From here it is easy to write down the expression for the new contribution to the lepton MDM, closely following~\cite{Ilisie:2015tra}:
\beq
\Delta a_\ell \simeq - 2\epsilon_\gamma \frac{m_\ell}{\Lambda_\gamma} \frac{\alpha Y'_\ell}{\pi} 
\log \left(\frac{\Lambda_\gamma}{M}\right) - 
2\tilde \epsilon_\gamma \frac{m_\ell}{\Lambda_\gamma} \frac{\alpha \tilde Y'_\ell}{\pi} 
\log \left(\frac{\Lambda_\gamma}{M}\right)
\eeq
where we have explicitly integrated out the heavy fermions.\footnote{From integrating out the heavy fermions
loop we have $\epsilon_\gamma = \frac{N_f Q_f^2 Y_f}{12 \pi^2}$ and 
$\tilde \epsilon_\gamma = \frac{N_f Q_f^2 \tilde Y_f}{8 \pi^2}$.  One can also match 
$\epsilon_g = \frac{N_f I_f Y_f}{12 \pi^2} $ and $\tilde \epsilon_g = \frac{N_f I_f \tilde Y_f}{8 \pi^2} $. In these
expressions $I_f = 1/2$ for the fundamentals of the $SU(3)$ and $I_f = 3$ for the adjoint.} 
In the case of $\ell=\mu$, a comparison with
the reported value of $a_\mu$ yields at the $2 \sigma$ level
\beqa
8.5 \times 10^{-3} & \lesssim |\epsilon_\gamma Y'_\mu| & 
 \lesssim 3.3 \times 10^{-2} \ \ \ \ {\rm for} \ \Lambda_\gamma = 1~{\rm TeV}
\eeqa 
Of course similar values apply for the combination $\tilde \epsilon_{\gamma } \tilde Y'_\mu$ if the 
$S$ is (predominantly) a pseudo-scalar.  These numbers should be taken with a grain of salt because of our approximation of the effects of the dimension-five operator by a simple logarithm. 
In particular, given that the expected separation of $M$ and $\Lambda_\gamma$ scales is expected to be moderate at most, the 
model dependent finite contributions could be comparable in specific UV-models.  

Given that $S \to \mu^+ \mu^-$ non-observation implies $|Y'_{\mu}| \lesssim 0.06$, 
we see that interpreting $\Delta a_\mu$ as a signal is marginally consistent with 
$\Lambda_\gamma$ being close to TeV. More importantly, the preferred range of $|Y'_\mu|$
will be probed in the very near future at the LHC13. Note also, that in order 
to interpret $\Delta a_\mu$ as a signal one is forced to assume $\epsilon_\gamma \gtrsim 0.1$. 
This assumption is in tension with non observation of $S$ in the di-photon final state at LHC8 
and suggests that the theory might have non-perturbative UV-completion~\cite{Franceschini:2015kwy}.
If instead interpreted as a  constraint, 
$a_\mu$ implies no bound which would be stronger than existing direct LHC searches in the 
$\mu^+ \mu^-$ channel. 

When we apply these considerations to $\epsilon_\gamma Y'_\tau$ and $\tilde \epsilon_{ \gamma}
\tilde Y'_{\tau}$, we get no meaningful bound, while bounds on 
the couplings of the new scalar to the electrons are marginal
\beq
|\epsilon_{\gamma} Y'_e|, |\tilde \epsilon_{\gamma} \tilde Y'_e| \lesssim {1.17 \times 10^{-2}\ \ \ \ {\rm for} 
\ \Lambda_\gamma = 1~{\rm TeV} }\,,
\eeq
which is comparable to the existing direct LHC constraints.

Turning to the leptonic EDMs, they constrain the CP-violating products of $S$ couplings 
 to the charged SM leptons and photons ($\epsilon_{\gamma}\tilde Y'_\ell$ and/or $\tilde \epsilon_{\gamma} Y'_\ell $). The current experimental bounds read 
\beqa
|d_e| <  8.7 \times 10^{-29}~{\rm e \ cm} \,,\\
d_\mu = (-0.1 \pm 0.9) \times 10^{-19}~{\rm e \ cm}\,, \\
(-0.25 < d_\tau < 0.008) \times 10^{-16}~{\rm e \ cm }\,.
\eeqa
We estimate the contributions of $S$ 
using the analogous expressions for the 
corresponding Barr-Zee diagrams, which were obtained 
for the hypothetic CP-violating Higgs couplings to the third generation fermions in~\cite{Brod:2013cka}.   

As in the MDM case, effectively integrating out the heavy fermions and keeping only the leading logarithmic
pieces we get the following expression for the new contribution to the lepton EDMs: 
\beq
\frac{d_\ell}{e} \simeq -\epsilon_\gamma\frac{\alpha }{ \pi} \frac{\tilde Y'_\ell}{\Lambda_\gamma}
\log\left(\frac{\Lambda_\gamma}{M}\right)~.
\eeq
Of course, similar contributions proportional to the $\tilde \epsilon_{\gamma} Y'_\ell$ will be formed. 
Evidently, no meaningful constraints exist on the CP-violating couplings either of the tau lepton or of the muon. However, 
the constraint on the CP-violating coupling to the electron is stringent
\beq
|\tilde \epsilon_{\gamma} Y'_e|, |\epsilon_{\gamma} \tilde Y'_e|   <   6.2 \times 10^{-9} \,,
\eeq
depending whether the contribution to the EDM is positive or negative (a priori, we do not know 
the sign of the products $\epsilon_\gamma \tilde Y'_e$, $\tilde \epsilon_{\gamma} Y'_e$). 

Finally we analyze the CP-violating couplings of  $S$ to the SM quarks. These are bounded 
by the measurements of nuclear EDMs, the most stringent constraints currently coming from the EDMs of the neutron and the nucleus of mercury.  A priori, there are several different
contributions to nuclear EDMs. Here we focus on quark EDMs, chromo EDMs (CEDMS), CP-violating nucleon-pion interactions and the contribution from the three-gluon Weinberg 
operator.
In our analysis we follow the procedure of~\cite{Kamenik:2011dk, Brod:2013cka}, 
defining the effective operators contributing the (C)EDMs as
\begin{align}
\mathcal L^q_{\slashed{CP}} &= \sum_q \left[ i \frac{c_q}{2} e Q_i \bar q \sigma^{\mu\nu} \gamma_5 q F_{\mu\nu} + i \frac{c^c_q}{2} g_s q \sigma^{\mu\nu} T^a \gamma_5 q G^{a}_{\mu\nu} + i C_q (\bar q q) (\bar q \gamma_5 q) \right] \nonumber\\
& + \frac{c_W }{6} g_s f^{abc} G^a_{\mu\rho} G^{b\rho}_{\nu} G^c_{\lambda\sigma} \epsilon^{\mu\nu\lambda\sigma}\,.
\end{align}
Analogously to the case of leptonic EDMs, the Wilson coefficients  of the EDM and CEDM operators at the matching scale (at the leading logarithmic order) are 
\beqa
c_q & \simeq & \epsilon_\gamma \frac{\alpha}{\pi}  \frac{\tilde Y'_q}{\Lambda_\gamma} 
\log \left(\frac{\Lambda_\gamma}{M}\right)\,, \\
 c^c_q & \simeq & \epsilon_g \frac{ \alpha_s}{\pi}  \frac{\tilde Y'_q}{\Lambda_\gamma} 
\log \left(\frac{\Lambda_\gamma}{M}\right)\,,
\eeqa 
and similarly for $\tilde \epsilon_{\gamma} Y'_q$, $\tilde \epsilon_{g} Y'_q$ combinations. We further run these operators to the hadronic scale $\mu_H\sim 1$~GeV using the known QCD 
expressions~\cite{Degrassi:2005zd, Hisano:2012cc}, including finite shifts of the Weinberg operator at heavy 
quark thresholds~\cite{Braaten:1990gq, Chang:1991ry} and match them onto the neutron and mercury 
EDMs (c.f.~\cite{Pospelov:2005pr}). Using the recently updated values of the relevant hadronic matrix 
elements~\cite{Bhattacharya:2015esa,Bhattacharya:2015wna} we obtain
\begin{align}
d_n & = 0.774(66) e Q_d c_d - 0.233(28) e Q_u c_u + 1.1(5) e (  c^c_d + 0.5  c^c_u) + 22(10) \times 10^{-3} {\rm GeV} e c_W \,,\\
d_{Hg} & = 7(4) \times 10^{-3} e ( c^c_u -  c^c_d) +  1.4 (7) \times 10^{-5} {\rm GeV}^2 e \left(\frac{C_{u}}{m_u} - 0.5 \frac{C_d}{m_d}\right)\,.
\end{align}
in terms of the effective Wilson coefficients evaluated at the hadronic scale. 
Interestingly the nuclear EDM bounds (at $90\%$ C.L.) of
 \beq
|d_n| < 2.9 \times 10^{-26}~e\ {\rm cm}\,, \ \ \ \ |d_{Hg}| < 3.1 \times 10^{-29}~e\ {\rm cm} \,,
\eeq 
 impose meaningful bounds on the CP-violating couplings.\footnote{In our analysis we do not include the bound on the CP-violating  coupling to the strange quark. The reason is that the neutron EDM dependence on this coupling 
is highly uncertain and consistent with zero~\cite{Fuyuto:2012yf,Bhattacharya:2015esa,Bhattacharya:2015wna}.} In particular, conservatively taking the lower one-sigma values for all hadronic matrix elements, the neutron EDM yields
\beqa
|\tilde Y'_d [(69\pm28)\epsilon_g - \epsilon_\gamma]| & < & 1.6 \times 10^{-5}\,,\\
|\tilde Y'_u [(60\pm24) \epsilon_g - \epsilon_\gamma]| & < & 2.6 \times 10^{-5}\,,\\
|\tilde Y'_c \epsilon_g| & < & 9.0 \times 10^{-4} \,,\\
|\tilde Y'_b \epsilon_g| & < & 9.6 \times 10^{-3} \,,\\
|\tilde Y'_t \epsilon_g| & < & 1.8\,,
\eeqa
while the mercury EDM gives 
\beqa
|\tilde Y'_d \epsilon_g| & < & 8.7 \times 10^{-4}\,,\\
|\tilde Y'_u \epsilon_g| & < & 8.7 \times 10^{-4}\,. 
\eeqa

In addition, the mercury EDM is also sensitive to CP violating four-fermion interactions of l
ight quarks~\cite{Pospelov:2005pr}, in particular tree level $S$ exchange yields
\beq
C_q = \frac{Y'_q \tilde Y_q}{M^2}\,,
\eeq
at the matching scale $M$. The QCD evolution to the hadronic scale~\cite{Hisano:2012cc} reduces $C_q$ by $C_q(\mu_H) \simeq 0.2 C_q(M)$ leading to the bounds 
\beqa
|\tilde Y'_d Y'_d | & \lesssim 8.9 \times 10^{-3}~, \\
|\tilde Y'_u Y'_u | & \lesssim 2.0 \times 10^{-3}~. 
\eeqa  

Finally, similar effects due to heavy quarks can be relevant though their contributions to the Weinberg operator. The two loop contribution due to a massive quark $q$ can be factorized in terms of the mixing of $C_q$ into $\tilde c_q$ followed by a threshold shift of $c_W$ at the scale $m_q$. This yields~\cite{Hisano:2012cc}
\beq
c_W \simeq  \frac{4 \alpha_s}{ (4\pi)^3} \frac{Y'_q \tilde Y'_q}{M^2}  \left[ \log \left( \frac{M}{m_q}\right) -\frac{3}{4}\right]~,
\eeq
in agreement with an explicit evaluation of the two-loop result~\cite{Dicus:1989va, Brod:2013cka}. In this case the bound coming from the neutron EDM is much stronger and yields
\beqa
|\tilde Y'_c Y'_c | & \lesssim &  1.0 \times 10^{-2}~. \\
|\tilde Y'_b Y'_b | & \lesssim& 4.3 \times 10^{-2}~, \\
|\tilde Y'_t Y'_t | & \lesssim& 1.3. 
\eeqa  

\subsection{Flavor observables}

Next we derive constraints on $Y_{i\neq j}$ and consequently flavored $\X$ decay products in Table~\ref{tab:1}. As before we restrict our discussion to the inclusion of a single non-vanishing $Y_{ij}$ at a time. Here the leading effects can appear already at the tree level. In particular, off-diagonal  couplings to quarks will induce neutral meson oscillations through tree-level $S$ exchange
\beq
\mathcal L_{\Delta F=2} = \sum_{i\neq j} \frac{Y_{d_i d_j}^{ 2}}{2M^2} (\bar d_i P_L d_j)^2 +  \sum_{i\neq j} \frac{Y_{u_i u_j}^{2}}{2M^2} (\bar u_i P_L u_j)^2 + \rm h.c.\,. \label{eq:DF=2}
\eeq
The resulting constraints can be derived following the results of~\cite{Blankenburg:2012ex,Harnik:2012pb}. The values shown in Table~\ref{tab:1} have been obtained taking into account the recent Lattice QCD updates~\cite{Bertone:2012cu,Carrasco:2013zta,Carrasco:2014uya}. We note that in case $S$ is a part of a larger $SU(2)_L$ multiplet there can be additional effects coming from the associated charged scalars. Formally at least, these will be loop suppressed compared to contributions in~\eqref{eq:DF=2}. In practice however the relative size of both effects depends on the detailed flavor composition of the model. 

The second class of effects is very similar to the contributions of flavor diagonal $Y_{ii}$ to the fermionic dipole moments discussed in the previous subsection. Namely, the simultaneous presence of $Y_{ij}$ and $\epsilon_{\gamma,g}$  generates radiative FCNC operators at the one-loop order. We define the relevant flavor changing dipole operators through
\beq
\mathcal L_{f_1,f_2} = c_{L f_1 f_2 \gamma} Q_{L f_1 f_2 \gamma} + (L \leftrightarrow R) + (\gamma \leftrightarrow g)\,,
\eeq
where
\beq
Q_{L f_1 f_2 \gamma} = \frac{e}{8 \pi^2} \bar f_2 (\sigma\cdot F) P_L f_1\,, \quad Q_{L f_1 f_2 g} = \frac{g_s}{8 \pi^2}  \bar f_2 T^a (\sigma\cdot G^a) P_L f_1\,.
\eeq
and analogous for $Q_{R f_1 f_2 \gamma} $ and $Q_{R f_1 f_2 g} $.  Using the results of existing calculations done in the context of renormalizable models~\cite{Davidson:2010xv,Goudelis:2011un,Harnik:2012pb} we derive
\begin{align}
c_{L f_1 f_2 \gamma} &\simeq { 4 \pi \, \alpha\, } \epsilon_\gamma  \, Q_{f_1}\,  \, \frac{Y_{f_2 f_1}}{ \Lambda_\gamma }  \log\left(\frac{\Lambda_\gamma}{M}\right)\,, \\
c_{L f_1 f_2 g} &\simeq { 4 \pi \, \alpha_s\, } \epsilon_g  \, \frac{ Y_{f_2 f_1}\, }{ \Lambda_g } \log\left(\frac{\Lambda_g}{M}\right)\,,
\end{align}
where we have again only kept the logarithmically enhanced contributions, and where $c_{R f_1 f_2 \gamma} = c^*_{L f_2 f_1 \gamma}$, $c_{R f_1 f_2 g} = c^*_{L f_2 f_1 g}$. Finally, contributions of $\tilde \epsilon_{\gamma}$ and $\tilde \epsilon_{g}$ can obtained directly from the above expressions via the replacements $\gamma \to \tilde \gamma$, $g \to \tilde g$ and $Y_{f_2 f_1} \to Y^*_{f_1 f_2}$\,. 
The partonic radiative rates are then given by
\begin{align}
\Gamma(f_1 \to f_2 \gamma) & =  \frac{\alpha m_{f_1}^3}{64 \pi^4} \left( |c_{L f_1 f_2 \gamma}|^2 + |c_{R f_1 f_2 \gamma}|^2 \right)\,, \\
\Gamma(f_1 \to f_2 g) & =  \frac{\alpha_s m_{f_1}^3}{48 \pi^4} \left( |c_{L f_1 f_2 g}|^2 + |c_{R f_1 f_2 g}|^2 \right)\,. 
\end{align}
We note in passing that similarly to the case of $\Delta F=2$ observables, in models where $\X$ is a member of a larger electroweak multiplet additional contributions can arise from one-loop exchange of the associated charged scalars. However, contrary to the  $\Delta F=2$ case, here such effects are not formally suppressed by either higher loop order or higher power in $1/\Lambda$ expansion. In fact it is well known, that within THDMs charged Higgs contributions typically provide the largest effects on radiative decays (c.f.~\cite{Misiak:2006ab}). In view of this, the bounds derived from dipole transitions can be viewed in such context as conservative estimates and stronger constraints are generically expected.

Imposing~\eqref{eq:GM} together with stringent bounds on radiative LFV decays of heavy leptons allows to derive strong constraints on all $Y_{\ell_i,\ell_j}$\,. On the other hand, due to existing $\Delta F=2$ bounds on $Y_{q_i q_j}$, radiative decays of hadrons do not generally impose additional constraints.  An exception are the  $b\to s\gamma$ and $b\to s g$ transitions in the $B$ sector, as well as $s\to d g$ transitions in the kaon sector, which we thus examine in detail.

\begin{figure}[!t]
\centering
\includegraphics[angle=0,width=7.cm]{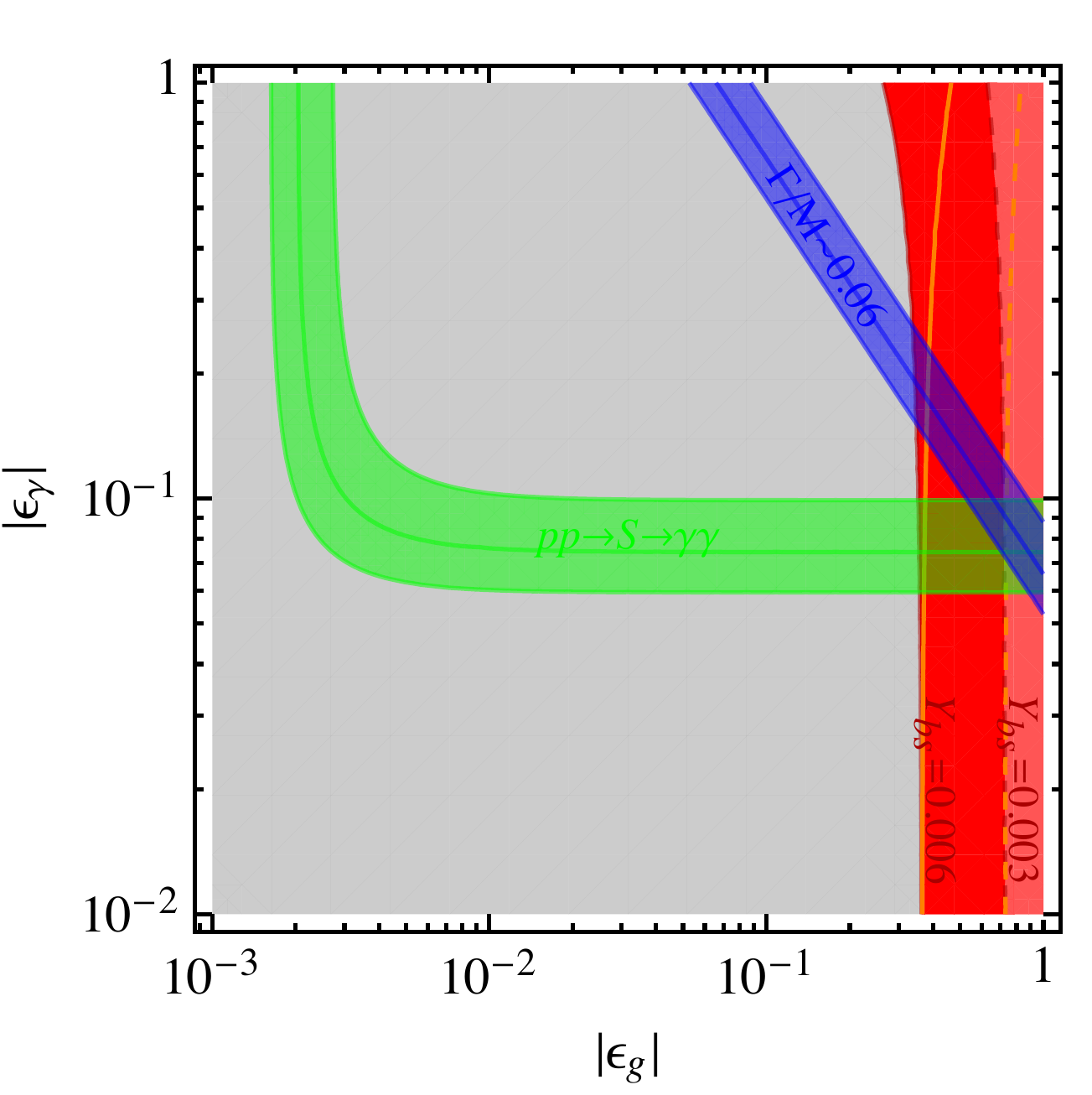}
\caption{\label{fig:bsgamma} Combination of bounds on the $\X$ couplings to photons and gluons coming from: the observation of $\X\to\gamma$ events at the LHC   (green band assuming gluon fusion dominated $S$ production),  the ATLAS inferred $\X$ resonance width (in blue band assuming a possible third decay channel saturating the width) and $b\to s \gamma$ constraints (in red bands) for various values of $\X$ couplings to $b\bar s$ ($Y_{bs}$)  allowed by $B_s$ oscillation measurements. The dark red  (orange) contours refer to positive (negative) relative sign between $\epsilon_\gamma$ and $\epsilon_g$. See text for details.}
\end{figure}

The $b\to s\gamma$ and $b\to s g$ transitions are severily constrained by the measurements of the inclusive $B\to X_s\gamma$ branching fraction and more generally, a global analysis of the inclusive and exclusive $b\to s \gamma$ and $b \to s \ell^+\ell^-$ FCNC mediated decays.   For the case of $B\to X_s\gamma$ one can write~\cite{Misiak:2015xwa}
\beq
\mathcal B(B\to X_s \gamma) \times 10^4 = (3.36\pm 0.23) - 8.22 {\rm Re}(\Delta C_7) - 1.99 {\rm Re} (\Delta C_8)\,,
\label{eq:bsg}
\eeq
where the terms quadratic in NP contributions and tiny effects from their imaginary parts have been neglected. $\Delta C_{7,8}$ are in turn defined as~\cite{Misiak:2006ab}
\beq
\Delta C_7 = \frac{c_{R b s \gamma}}{\sqrt{2} m_b G_F V_{tb}^*V_{ts}}\,,\quad \Delta C_8 = \frac{c_{R b s g}}{\sqrt{2} m_b G_F V_{tb}^*V_{ts}}\,.
\eeq
Equation~\eqref{eq:bsg} holds at the renormalization scale $\mu_0 = 160$~GeV. In our analysis we neglect small QCD running effects from our effective field theory (EFT) cut-off scale $\Lambda_i=1$~TeV. The teoretical prediction in~\eqref{eq:bsg} has to be compared to the current experimental world average of $\mathcal B(B\to X_s \gamma)_{\rm exp} = (3.43 \pm 0.21 \pm 0.07) \times 10^{-4}$~\cite{Amhis:2014hma}. Combined with the bounds from $B_s$ oscillations and~\eqref{eq:GM}, this leads to the 2$\sigma$ constraints shown in Fig.~\ref{fig:bsgamma}. 
We observe that if $Y_{bs}$ is close to the bound set by $B_s$ oscillation measurements, a large $\X$ width in~\eqref{eq:Gamma} cannot be reproduced with only the $\X\to gg$ mode.

On the other hand $c_{L b s \gamma}$ and $c_{L b s g}$ are most stringently constrained by a global fit to $b \to s \gamma$ and $b \to s \ell^+\ell^-$ related measurements (c.f.~\cite{DescotesGenon:2011yn}). In particular one obtains
\beq
C'_7 = \frac{c_{L b s \gamma}}{\sqrt{2} m_b G_F V_{tb}^*V_{ts}}\,,\quad C'_8 = \frac{c_{L b s g}}{\sqrt{2} m_b G_F V_{tb}^*V_{ts}}\,,
\eeq
where a recent analysis~\cite{Altmannshofer:2014rta} assuming real $C'_{7,8}$ yields a constraint on the combination
\beq
-0.1 < \eta_{sb}\, {\rm Re}(C'_7 + 0.24 C'_8) < 0.12\quad \rm at~ 95\% ~C.L.,
\eeq
 where $\eta_{sb} \simeq 0.58$ accounts for the QCD running between $\mu_0$ and the $b$-quark mass scale~\cite{DescotesGenon:2011yn}\,. 
 In this case the constraints are milder and any value of $Y_{sb}$ below the bound set by $B_s$ oscillation measurements does not significantly constrain the phenomenology of $\X$\,.

In the kaon sector, dipole operators mediating $s\to d g$ transitions can be probed via the measurements of direct CP violation in $K^0 \to 2\pi$ decays, in particular $\epsilon'/\epsilon$. Interestingly, the most recent SM prediction for this quantity~\cite{Buras:2015yba}
\beq
(\epsilon'/\epsilon)_{\rm SM} = (1.9\pm 4.5) \times 10^{-4}\,,
\eeq
exhibits a $\sim 2\sigma$ tension with the corresponding experimental measurements~\cite{Batley:2002gn, AlaviHarati:2002ye, Worcester:2009qt} yielding
\beq
(\epsilon'/\epsilon)_{\rm exp} = (16.6\pm 2.3) \times 10^{-4}\,.
\eeq
In turn, loop induced $S$ contributions to $c_{Lsdg}, c_{Ldsg}$ could  play  a  role in bridging  the  gap  between experiment and the SM expectations. The relevant effect in $\epsilon'/\epsilon$ can be written as (c.f.~\cite{Buras:2015yba})
\beq
\left|{\epsilon'}/{\epsilon}\right|_{c_{Lsdg}} = \frac{w}{\sqrt{2} |\epsilon_K| {\rm Re}A_0} \left| {\rm Im}A_0\right|_{c_{Lsdg}} \,, \quad \left| {\rm Im}A_0\right|_{c_{Lsdg}} = |\eta_{sd} {\rm Im} (c_{Lsdg}) \langle (2\pi)_{I=0} | Q_{Lsdg} | K^0\rangle|\,,
\eeq
and analogously for $c_{Ldsg}$. In our numerical evaluation we employ $1/w=22.46$, $|\epsilon_K| = 1.596 \times 10^{-3}$, ${\rm Re}A_0 = 33.22 \times 10^{-8}$~GeV, while $\eta_{sd}=0.88$ accounts for the QCD running of $c_{Lsdg}$ from the matching scale to $m_c$~\cite{Buras:1999da}. The value of relevant matrix element is still fairly uncertain. A recent estimate~\cite{Buras:2015yba}  using a  chiral quark model gives
\beq
\langle (2\pi)_{I=0} | Q_{Lsdg} | K^0\rangle \simeq [2-8]\times 10^{-2} {\rm GeV}^2\,.
\eeq
Conservatively using the lower boundary of this interval and combining the remaining theoretical and experimental uncertainties in quadrature, we obtain a $2\sigma$ preferred range of $\epsilon_g {\rm Im} Y_{sd}$ of 
\beq
1.4 \times 10^{-6} <  |\epsilon_g {\rm Im} Y_{sd}| < 7.5 \times 10^{-6}\,.
\eeq
Notice that such values can be consistent with constraints coming from neutral kaon oscillations provided $|\epsilon_g| \gtrsim 0.01$\,. Analogous conclusions hold for the product $\tilde \epsilon_g {\rm Re} Y_{sd}$ as well as $\tilde \epsilon_g {\rm Re} Y_{ds}$ or $ \epsilon_g {\rm Im} Y_{ds}$.

Finally, we note that all couplings of $\X$ to the top quark are only very weakly constrained. In particular radiative $t \to q \gamma$ and $t\to q g$ decays do not impose any relevant constraints on $Y_{tq}$, $Y_{qt}$\,. More information can in fact be extracted from existing LHC flavor tagged analyses at high $p_T$. While there are not current searches for singly produced $tj$ resonances, we can recast the recent ATLAS searches~\cite{Aad:2014xea,Aad:2015typ} for $t\bar b$ resonances taking into account the poorer efficiency for $b$-tagging charmed jets ($0.2$ versus $0.7$ for $b$-jets) but neglecting possible differences in acceptance and efficiency due to different production ($gg$ versus $\bar q q'$). In the narrow width approximation, which applies for  $\Gamma/M \ll 0.1$ the bound is
\beq
\frac{\Gamma(S\to t \bar c)}{\Gamma(S\to \gamma\gamma)} \lesssim 1500 \times \left( \frac{6\rm fb}{\sigma(pp\to S)_{13\rm TeV} \mathcal B(S\to \gamma\gamma)}\right) \left( \frac{R_{\rm 13TeV/8TeV}}{5}\right)\,,
\eeq
where $R_{\rm 13TeV/8TeV} \equiv \sigma(pp \to S)_{13\rm TeV}/\sigma(pp \to S)_{8\rm TeV}$. On the other hand currently no relevant experimental bound on $\Gamma(S\to t \bar u)$ can be extracted in this way\,.

\subsection{Implications of $\bf S \to \bf \gamma\gamma$ for  $\bf R_K$ puzzle}

Recently, the LHCb collaboration has reported discrepancies  in angular observables in the rare $b\to s \mu^+ \mu^-$ FCNC mediated decay $B\to K^* \mu^+ \mu^-$~\cite{Aaij:2013qta, LHCb:2015dla} as well as in the ratio of decay rates of $B^+\to K^+ \mu^+ \mu^-$ versus $B^+\to K^+ e^+ e^-$~\cite{Aaij:2014ora}. Interestingly, both anomalies can be successfully accommodated by supplementing the SM contributions with  
\beq
\mathcal H_{\rm eff}=  [\bar s \gamma^\mu (c P_L + c' P_R) b]  [\bar \mu  \gamma_\mu (v + a \gamma_5)  \mu]\,,
\eeq
where a good fit to all relevant $b\to s \mu^+ \mu^-$ data can be obtained for several choices of non-vanishing parameters~\cite{Altmannshofer:2014rta,Altmannshofer:2015sma}
\beq
[c v \simeq (34{\rm TeV})^{-2} ] \quad {\rm or} \quad [c v = -c a \simeq (48 {\rm TeV})^{-2} ] \quad {\rm or} \quad [c' v = - c v \simeq (35 {\rm TeV})^{-2} ]\,.
\eeq
Such contributions cannot be generated via tree-level exchange of a neutral scalar. Instead, perhaps the simplest explanation is due the exchange of a neutral massive vector $U(1)'$ gauge boson ($Z'$)~\cite{Gauld:2013qba, Altmannshofer:2014cfa}.  Interestingly such models, if weakly coupled, require the introduction of scalars neutral under SM gauge group as well as heavy vector-like quark and lepton partners. The scalar condensate controls both the spontaneous breaking of the $U(1)'$ as well as the mixing of SM fermions with their vector-like counterparts, which endows them with couplings to the $Z'$. 

To examine the possibility that $\X$ represents the physical excitation of the $U(1)'$ breaking condensate we consider the following simple model (see~\cite{Altmannshofer:2014cfa, Niehoff:2015bfa} for related previous work). We introduce two pairs of vector-like fermions $\mathcal Q$ and $\Psi$ with SM and $U(1)'$ gauge quantum numbers $(3,2,1/6,Q')$ and $(1,2,-1,Q')$ as well as $\mathcal X = (w + \X)/\sqrt{2}$ carrying $(1,1,0,Q')$ and with $w \equiv \sqrt{2} \langle \mathcal X \rangle$. All SM fields are assumed to be neutral under $U(1)'$. For any nonvanishing $U(1)'$ charge $Q'$, the only terms coupling SM fermions to the new sector are the $\mathcal X$ Yukawa couplings between $\mathcal Q$, $\Psi$ and the corresponding SM fermion doublets $q$ and $\ell$ 
\beq
\mathcal L_{\rm mix} = - \tilde Y_{q} \bar{\mathcal Q} \mathcal X q - \tilde Y_{\ell} \bar{\Psi} \mathcal X \ell\,.
\eeq
If they mix with a single SM generation the weak and mass eigenstates will be related via~\cite{Dawson:2012di,Fajfer:2013wca}
\beq
\left(\begin{array}{c} q \\ \mathcal Q \end{array}\right) = \left(\begin{array}{cc} \cos\theta & \sin\theta_{q} \\  -\sin\theta_q & \cos\theta_q \end{array} \right) \left(\begin{array}{c} q' \\ \mathcal Q' \end{array}\right)
\eeq
where
\beq
\tan 2\theta_q = \frac{\sqrt{2} M_{\mathcal Q} w \tilde Y_{q}}{M_{\mathcal Q}^2 - (w \tilde Y_{q})^2/2 - (v y_{q})^2/2}\,,
\eeq
where $M_{\mathcal Q}$ is the $\mathcal Q$ Dirac mass, $y_q$ is the SM fermion Yukawa coupling to the Higgs and $v=246$~GeV. The mass eigenstates are then determined via 
\beq
m_{\mathcal Q} m_q = M_{\mathcal Q} y_q \frac{v}{\sqrt 2}\,, \quad m_q^2 + m_{\mathcal Q}^2 = M_{\mathcal Q}^2 + \frac{w^2 }{2} \tilde Y_q^2 + \frac{v^2}{2} y_q^2\,.
\eeq
The analogous expressions for leptons can be obtained via the replacement $q \to \ell$.  Note that due to the different up- and down-type quark masses within the SM weak doublet, $\theta_q$ and $m_{\mathcal Q}$ can differ between the up and down sectors. In practice however the differences are negligible for $M_{\mathcal Q}\gtrsim 1$~TeV. Secondly, since the SM fermion mixes with a state with exactly the same SM quantum numbers, modifications of $W$ and $Z$ couplings to the physical state $q$ are protected by both EW and $U(1)'$ breaking and thus safely small. Together, both effects make this model largely transparent to EW precision tests. Consequently even maximal mixing is perfectly allowed. It does however significantly affect $U(1)'$ and $\X$ interactions. In particular, mixing induces perturbative ($\mathcal Q$, $q$) loop contributions to radiative $\X$ decays. Secondly, it provides new decay channels for $\X$: for example if $\Psi$ is lighter than $\X$, the later can decay to a $\Psi \ell$ pair. Such decay rates are proportional to $(\tilde Y_\ell \cos \theta_\ell)^2$\,. 

Finally, the induced $U(1)'$ charges of $q$ and $\mathcal \ell$ can have effects in low energy processes of SM fermions. In particular, if we furthermore assume, that $\mathcal Q$ only mixes with the third generation SM doublet in the up-quark mass basis, where $q = (t_L, \sum_i V_{ti} d^i_L)^T$, while $\Psi$ only mixes with the second generation SM doublet in the charged lepton mass basis, where $\ell = (\sum_i U_{i\mu} \nu_i, \mu_L)^T$ we obtain 
\beq
cv = -ca = \frac{\sin^2\theta_q \sin^2\theta_\ell {|V_{tb}^*V_{ts}|}}{w^2}\,.
\eeq
We see that for $w\lesssim 1$~TeV a preferred set of values for $cv,ca$ can in principle be reproduced. However, the same interactions also generate contributions to $B_s$ mixing. In particular~\cite{Altmannshofer:2014cfa}
\beq
\frac{M_{12}}{M_{12}^{\rm SM}} = 1+ \Lambda_{\rm SM}^2  \sin^4 \theta_q (V^*_{tb}V_{ts})^2 \left( \frac{1}{w^2} + \frac{1}{16\pi^2} \frac{1}{M_{\mathcal Q}^2} \right)\,,
\eeq
where $\Lambda_{\rm SM} = 4 \pi m_{W} / g_W V^*_{tb} V_{ts} \sqrt{S_0}  \simeq 40{\rm TeV}$\,. For $M_{\mathcal Q} \gtrsim 1$~TeV, the second term in the brackets is negligible. The same effect is present in $B_d$ mixing, by replacing $V_{ts} \to V_{td}$. Requiring $|{M_{12}}/{M_{12}^{\rm SM}}  -1 | < 0.15$ thus constrains $w/\sin^2 \theta_q \gtrsim 4$~TeV. This condition can be easily satisfied for even for $\tilde Y_q\sim 1$ at $M_{\mathcal Q} \sim 1$~TeV provided a low enough $U(1)'$ breaking  $w\lesssim 0.5$~TeV. Note that for $M=750$~GeV, this corresponds to a  $\mathcal X$ with a quartic value of $\sim 1$. 

On the other hand neutrino trident production constrains the $Z'$ couplings to muons. In our case, the relevant cross-section, normalized to the SM value is given by~\cite{Belusevic:1987cw} 
\beq
\frac{\sigma}{\sigma_{\rm SM}} \simeq \frac{(1+ 2v^2\sin^2\theta_\ell/w^2)^2+(1+4s_W^2 + 2v^2\sin^2\theta_\ell/w^2)}{1+(1+4s_W^2)^2}\,,
\eeq
where $s_W\simeq\sqrt{0.23}$ is the sine of the weak mixing angle. Comparing this with a recently compiled average of measurements $\sigma_{\rm exp}/\sigma_{\rm SM} = 0.83 \pm 0.18$~\cite{Altmannshofer:2014cfa} we obtain a $2\sigma$ bound on $w/\sin\theta_\ell \gtrsim 0.91$~TeV.\footnote{This bound makes the $Z'$ contribution to $a_{\mu}$ negligible in this model.}

\begin{figure}[!t]
\centering
\includegraphics[angle=0,width=8.2cm]{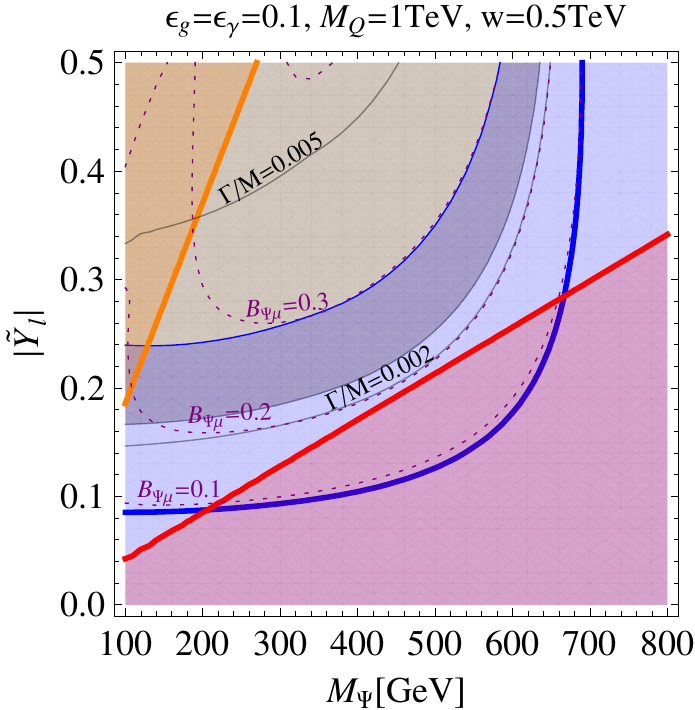}
\caption{\label{fig:singlet} Constraints on the parameter space of the gauged $U(1)'$ model addressing the observed LHCb anomalies in $B\to K^{(*)}\mu^+\mu^-$ decays: Hints of $\X$ production and decay to two photons at 13TeV LHC (central value in thick blue line, blue band allows for 50\% variation in signal strength), monojet bounds at 8TeV LHC (excluded region in brown), $B_s-\bar B_s$ oscillation measurements (excluded region in red), neutrino trident production (excluded region in orange). Overlaid are contours of constant $\Gamma/M$ (in full black lines) and $\mathcal B_{\Psi\mu} \equiv \mathcal B(\X \to \Psi^{\pm} \mu^{\mp})$ (in purple dotted lines)\,. See text for details.}
\end{figure}

The results of a combined fit to $\X$ signal strength and flavor data is shown in Fig.~\ref{fig:singlet}, where we also fix $\epsilon_\gamma=\epsilon_g = 0.1$ to reproduce the required radiative $\X$ widths to within 50\% (shown in blue shading). In addition requiring to reproduce $b\to s \mu^+ \mu^-$ data, the constraint coming from $B_s$ oscillation measurements and neutrino trident production exclude the red and orange shaded regions at small and large $\theta_\ell$, respectively.  
The $\X$ width is dominated by the decays $\X \to gg$, however at small $M_\Psi$ also $\X \to \mu^\pm \Psi^\mp$ and $\X \to \nu \Psi^0$ can become important. The invisible decay mode provides a constraint on this setup~\cite{Franceschini:2015kwy} (shown in brown shading in the figure). Consequently the ATLAS indicated large $\X$ width cannot be obtained with typical allowed widths of the order of $\Gamma/M \lesssim 0.002$. 
The charged $\Psi$ state is radiatively split in mass from the neutral component~\cite{Cirelli:2005uq, DelNobile:2009st, DiLuzio:2015oha} and will decay promptly as $\Psi^\pm \to \pi^\pm \Psi^0$. Since the pion in the decay is very soft and can be missed, the resulting signature for $\X \to \mu \Psi$ resembles a monolepton, but with an additional soft pion. 

Since low energy flavor observables do not directly depend on the $U(1)'$ gauge coupling, the mass of the corresponding gauge boson is undetermined. If it is light enough $\X$ can decay to it, providing additional interesting decay modes $\X \to 2 A' \to 4 j_b$ and $\X \to 2 A' \to 4 \mu$\,. Finally, in the narrow window where $M > 2 m_{A'} > 2 M_\Psi$, also $\X \to 2 A' \to 2\Psi^\pm 2\mu$ decay mode is possible, while $\X \to 2 A' \to 2\Psi^0 2\nu$ just contributes to the invisible $\X$ width.

On the other hand, the new vector-like fermions in the model can be searched for directly at colliders, such as the LHC. Limits on corresponding vector-like quarks are currently at ${\cal O}(800-900)$\,GeV \cite{Aad:2014efa,Aad:2015mba,Aad:2015gdg,Aad:2015kqa,Aad:2015voa,Khachatryan:2015gza,Khachatryan:2015oba}, while the prospects for the high-luminosity LHC (HL-LHC) at 14\,TeV are around 1500\,GeV \cite{Gershtein:2013iqa,Agashe:2013hma}. Turning to the lepton sector, constraints reported in the PDG stem so far from LEP, and are at $m_L \gtrsim 100$\, GeV~\cite{Agashe:2014kda}. Regarding the LHC, existing multi-lepton searches using the full $\sqrt s=8\,$TeV data set can be used to place limits on masses of vector-like leptons. These are rather model dependent and currently reside at around ${\cal O}(200-400)$\,GeV \cite{Falkowski:2013jya}. In turn, the some of the interesting parameter space of the model should be accessible with the next LHC run and finally at the HL-LHC.

\section{Conclusions}
\label{sec:final}

We have explored the potential physics underlying the excess in searches for di-photon resonances at ATLAS and CMS, parameterizing it in terms of higher dimensional operators including the SM fields and a new (pseudo-)scalar particle $S$.\footnote{While we focused on the case of $S$ being an SM singlet, we also commented on the electroweak-doublet hypothesis.} Beyond that, we considered renormalizable interactions of the latter with new vector-like fermions.
In particular, we have addressed the question of calculability in scenarios where the tentative signal
is mediated by loops of such new fermions, indicating that existing (apparent) explanations featuring sizable $S$ Yukawa couplings and a bunch of new fermions are in fact not calculable in perturbation theory. In particular, we have proposed a simple strategy to test if in a given model perturbation theory can be trusted. We applied this to a toy model of new vector-like fermions and derived concrete bounds on their number and their Yukawa couplings, confronting them with the parameter space of possible explanations of the anomaly.

Moreover, we have studied correlations between the di-photon excess and low energy observables.
A first focus was on dipole moments of SM fermions coupled to $S$, which are generated due to the (sizable) coupling of $S$ to photons (and gluons).
We showed that the constraints in this sector are able to shed light on the interactions of $S$. In particular, we demonstrated that an explanation of the muon MDM anomaly requires a $S \mu^+ \mu^-$ coupling of ${\cal O}(0.1)$ and can be tested by limits on the decay $S \to \mu^+ \mu^-$. The latter are already cutting significantly into the parameter space and will explore a large portion of it with the next LHC run. On the other hand, upper bounds on $a_e$  as well as on the corresponding electric dipole moment can significantly constrain the couplings of $S$ to electrons.

After that, we turned to correlations between the di-photon excess and flavor physics, considering four-fermion operators generated after integrating out $S$ at the tree level as well as radiatively generated dipole operators due to the $S \gamma \gamma$ and $S g g$ interactions. We consequently presented limits on the $S$ Yukawa couplings from measurements in the flavor sector, such as meson oscillations as well as FCNC decays of hadrons and leptons.
While in the quark case, usually limits from meson mixing provide the most stringent constraints,
we showed in particular that in the presence of new right-handed contributions to dipole operators,
bounds on $b \to s \gamma$ transitions, together with the requirement to reproduce the correct $S$ width, can add non-trivial information on the Yukawa couplings. We explored these constraints, together with those from the cross section of the $S\to \gamma \gamma$ signal and from $B_s$ oscillations in the $\epsilon_\gamma,\epsilon_g,Y_{sb}/Y_{bs}$ parameter space. Similarly interesting interplay occurs in the kaon sector, where we have shown that $\epsilon_g, Y_{sd}/Y_{ds}$ effects consistent with bounds from neutral kaon oscillation measurements could ameliorate the tension between the recent SM predictions and experimental measurements of the direct CP violation in $K^0 \to 2\pi$ decays ($\epsilon'/\epsilon$). 

Finally, we also considered the flavor anomalies recent found in the angular analysis of $B \to K^\ast \mu^+ \mu^-$ and in $R_K$. We showed that the scalar $S$ could be linked to the breaking of a gauged $U(1)^\prime$ symmetry, which in turn could address these anomalies via $Z^\prime$ exchange. We assumed the latter being coupled to new vector-like fermions, charged under the $U(1)^\prime$, that mix with the ($U(1)^\prime$ neutral) SM fermions only via the vacuum expectation value of a scalar that features $S$ as a physical excitation. We considered various constraints on this model, summarized in Figure \ref{fig:singlet}.
\vspace{0.4cm}

\paragraph{Acknowledgements}
We thank Luca Di Luzio for useful discussions. Special thanks goes also to Ulrich Haisch for his several enlightening comments on the first version of the manuscript.
JFK acknowledges the financial support from the Slovenian Research Agency (research core funding No. P1-0035).
FG acknowledges support by a Marie Curie Intra European Fellowship within the EU FP7 (grant no. PIEF--GA--2013--628224). 
MN acknowledges support of the STFC grant ST/L000385/1, and King's College, Cambridge.
\\

\noindent 
{\bf Note added:} Our paper appeared on arXiv on the same day as~\cite{Son:2015vfl}. 
This reference also analyzes calculability and perturbativity constraints on possible explanations of the 
750~GeV diphoton excess. 
Note, however, that our paper uses 
a different criterion for calculability.

\bibliography{refs}
\bibliographystyle{jhep}
\end{document}